\documentclass[journal]{IEEEtai}
\usepackage[colorlinks,urlcolor=blue,linkcolor=blue,citecolor=blue]{hyperref}

\usepackage{amsmath,amsfonts}
\usepackage{algorithmic}
\usepackage{algorithm}
\usepackage{array}
\usepackage[caption=false,font=normalsize,labelfont=sf,textfont=sf]{subfig}
\usepackage{stfloats}
\usepackage{url}
\usepackage{verbatim}
\usepackage{graphicx}
\usepackage{cite}
\usepackage{multirow}
\usepackage{lipsum}
\usepackage{booktabs}
\usepackage{textcomp}
\usepackage{xcolor}

\begin{document}

\title{NeuroAMP: A Novel End-to-end General Purpose Deep Neural Amplifier for Personalized Hearing Aids}

\author{Shafique Ahmed, Ryandhimas E. Zezario~\IEEEmembership{IEEE Member}, Hui-Guan Yuan, Amir Hussain~\IEEEmembership{IEEE Senior Member}, \\ Hsin-Min Wang ~\IEEEmembership{IEEE Senior Member}, Wei-Ho Chung, Yu Tsao~\IEEEmembership{IEEE Senior Member}}


\markboth{Journal of IEEE Transactions on Artificial Intelligence}
{First A. Author \MakeLowercase{\textit{et al.}}: Bare Demo of IEEEtai.cls for IEEE Journals of IEEE Transactions on Artificial Intelligence}

\maketitle

\begin{abstract}

The prevalence of hearing aids is increasing. However, optimizing their amplification remains challenging due to the complexity of integrating multiple components in traditional methods. To address this, we present NeuroAMP, a novel deep neural network for end-to-end, personalized amplification in hearing aids. NeuroAMP leverages spectral features and the listener's audiogram as inputs, and we explore four architectures: Convolutional Neural Network (CNN), Long Short-Term Memory (LSTM), Convolutional Recurrent Neural Network (CRNN), and Transformer. We also introduce Denoising NeuroAMP, an extension that integrates noise reduction with amplification for improved real-world performance. To enhance generalization, we employed a comprehensive data augmentation strategy during training on diverse speech (TIMIT, TMHINT) and music (Cadenza Challenge MUSIC) datasets. Evaluation using the Hearing Aid Speech Perception Index (HASPI), Hearing Aid Speech Quality Index (HASQI), and Hearing Aid Audio Quality Index (HAAQI) shows that the Transformer-based NeuroAMP achieves the best performance, with SRCC scores of 0.9927 (HASQI) and 0.9905 (HASPI) on TIMIT, and 0.9738 (HAAQI) on Cadenza dataset. Notably, the augmentation strategy maintains robust performance on unseen datasets (e.g., VoiceBank-DEMAND, MUSDB18-HQ). Furthermore, Denoising NeuroAMP outperforms both the conventional NAL-R+WDRC method and a two-stage baseline on the VoiceBank-DEMAND dataset, achieving HASPI of 0.90 and HASQI of 0.59. These results highlight the strong potential of NeuroAMP and Denoising NeuroAMP to provide a novel and effective framework for personalized hearing aid amplification.

\end{abstract}


\begin{IEEEImpStatement}
This study presents NeuroAMP, a novel deep learning-based hearing aid amplifier that replaces complex multi-stage processing with streamlined, end-to-end personalized amplification. We also introduce Denoising NeuroAMP, which integrates noise reduction into the amplification process. Validated on diverse datasets, these models outperform conventional and two-stage methods on standardized metrics such as HASPI and HASQI. Moreover, we believe NeuroAMP’s data-driven approach can advance hearing aid technology by enhancing user-centered assistance and supporting more effective communication for individuals with hearing loss. Future work will explore real-time implementation and incorporate user feedback to further refine personalization.
\end{IEEEImpStatement}


\section{Introduction}

\IEEEPARstart{H}{earing} loss, a pervasive global health issue, continues to impact the lives of hundreds of millions, leading to social isolation~\cite{shukla2020hearing}, depression~\cite{lawrence2020hearing}, diminished quality of life~\cite{ciorba2012impact}, and even cognitive decline~\cite{lin2013hearing}. The World Health Organization (WHO) highlights that approximately 430 million of people worldwide currently face problems related to hearing loss~\cite{WHO_hearing_loss_2024, Audiologists_org_2024}. Despite the availability of hearing aids, which can mitigate these adverse effects, only 16\% of individuals with hearing loss in the United States consistently use them~\cite{HA2018use, HA2018use2}. This low adoption rate suggests a notable gap between the potential of hearing aids and their real-world utilization, emphasizing the importance of developing more effective and user-friendly solutions.

\begin{figure}[t]
\centering
\includegraphics[scale=0.45]{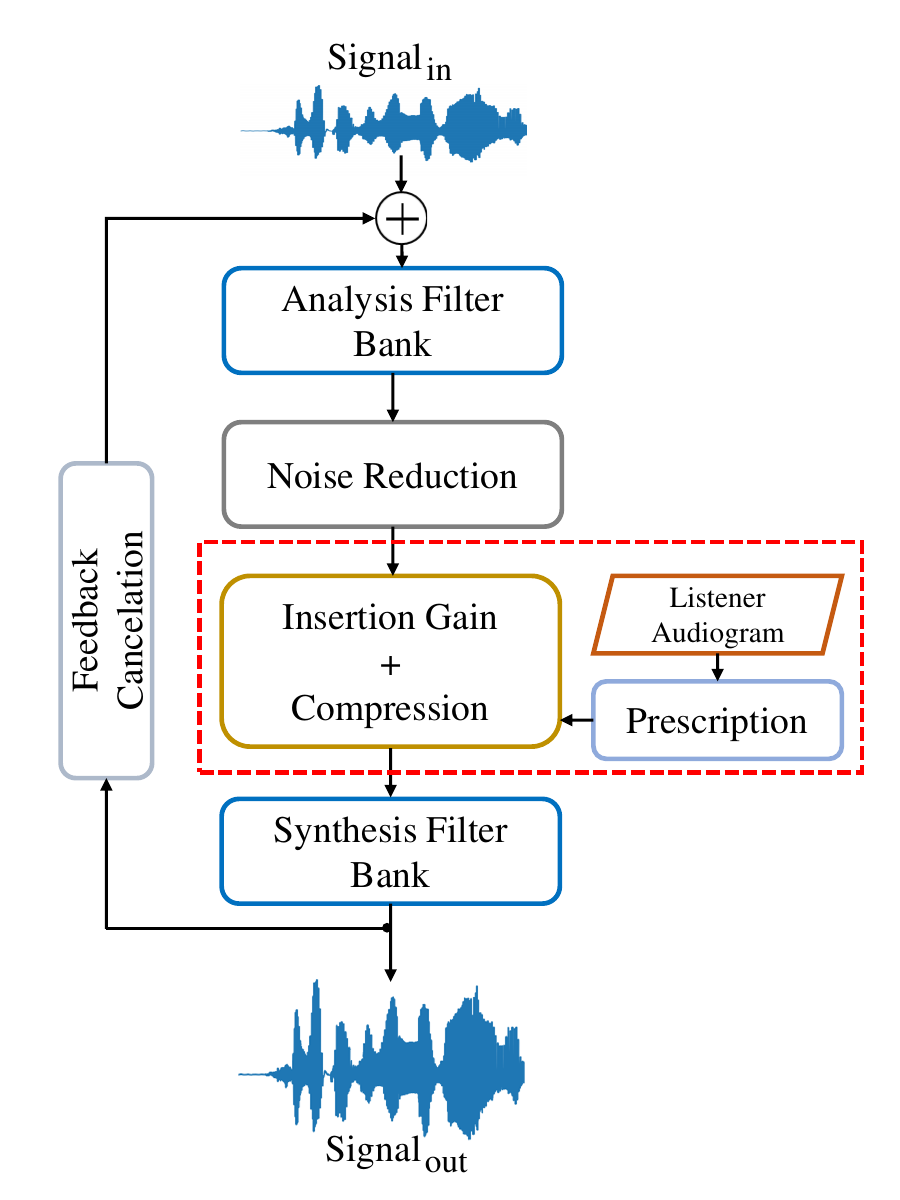}
\caption{Overview of a typical hearing aid system, illustrating the modular components involved in signal processing, including microphone input, insertion gain, compression, and speaker output.}
\label{fig_1}
\end{figure}

Non-use of hearing aids by people with hearing loss can be attributed to a variety of factors, including hearing aids being ineffective in noisy environments, suboptimal sound quality, insufficient benefit, and incompatibility with the individual's specific type of hearing loss \cite{HA2015use3}. To address these challenges, various amplifier strategies for hearing aids have been developed \cite{NAL-R, DSL, NAL-NL1, Compression, WDRCbenefit1}. These strategies primarily involve two key components: a prescription fitting formula, which determines the desired insertion gain from the individual's audiogram, and compression techniques, which are subsequently applied to manage the dynamic range of sounds, as shown in Fig.\ref{fig_1}. The well-known prescription fitting formulas include NAL-NL1\cite{NAL-NL1}, NAL-NL2~\cite{NAL-NL2}, NAL-R~\cite{NAL-R}, and DSL m[i/o]~\cite{DSLm}.

A common problem for people with hearing loss is a reduction in the auditory dynamic range, known as loudness recruitment. In this case, weak sounds may become inaudible, while loud sounds can be uncomfortably intense. Non-linear prescription formulas like NAL-NL1, NAL-NL2, and DSL m[i/o] play a crucial role in addressing the reduced auditory dynamic range through compression. These advanced formulas, however, are primarily used in commercial hearing aids and require licensing, which may limit their broader accessibility. In this study, we use the widely adopted NAL-R prescription formula and incorporate it with Wide Dynamic Range Compression (WDRC)~\cite{WDRCbenefit1} to introduce non-linearity, mimicking the behavior of more advanced systems like NAL-NL1. This approach aligns with the established practice in the open-source OpenMHA platform~\cite{OpenMHA}. 

Although conventional amplifiers demonstrate notable performance, they often face challenges in adapting to dynamic acoustic environments. These limitations in adaptability and optimization have led to the exploration of alternative approaches. One such promising approach is deep learning, which offers the potential to learn complex patterns from large datasets and create more adaptable systems. Recently, with the availability of huge amounts of training data and their corresponding labels, deep learning models have achieved notable performance in several speech processing applications. Additionally, the integration of deep learning models for speech processing in hearing aid devices has demonstrated robust performance, for example, in speech enhancement~\cite{wang2017deep,akeroyd20232nd} and speech assessment tasks~\cite{tu22_interspeech, edozezario22_interspeech, MAWALIM2023109663}.

In this paper, we propose a deep learning-based neural amplifier model called NeuroAMP, designed to provide personalized amplification in an end-to-end manner. Our initial goal is to demonstrate that NeuroAMP can achieve comparable performance to established methods, while also providing the potential for future improvements through its inherent adaptability. Unlike conventional amplifiers (as indicated by the red dashed box in Fig.~\ref{fig_1}) that combine prescription fitting formulas and compression techniques to amplify audio inputs, NeuroAMP uses a neural network to perform amplification in an end-to-end manner. We hypothesized that the non-linearity of the neural network would allow NeuroAMP to capture a greater variety of acoustic information and more effectively adjust gain and compression based on the user's specific degree of hearing loss at different frequencies. The development of NeuroAMP involved exploring various neural network architectures, including Convolutional Neural Networks (CNN), Long Short-Term Memory (LSTM), Convolutional Recurrent Neural Networks (CRNN), and Transformer. The training objective of NeuroAMP is to minimize the loss between the estimated amplified audio and the corresponding ground-truth (audio signal processed by conventional insertion gain and compression). To ensure robust performance across diverse conditions, we also apply a data-augmentation strategy using datasets containing speech from multiple languages as well as music. Experimental results confirm the superiority of the Transformer model over other models, achieving the best performance in almost all metrics (with Spearman's Rank Correlation Coefficient (SRCC) scores of 0.9927 for Hearing Aid Speech Quality Index (HASQI) and 0.9905 for Hearing Aid Speech Perception Index (HASPI) on the TIMIT dataset~\cite{timit}, and 0.9738 for Hearing Aid Audio Quality Index (HAAQI) on the Cadenza challenge dataset~\cite{cedenza1}). Additionally, we also verify the effectiveness of data augmentation in maintaining good performance on unseen datasets (with SRCC scores of 0.9810 for HASQI and 0.9914 for HASPI on the VoiceBank-DEMAND dataset~\cite{vctk}, and 0.9892 for HAAQI on the MUSDB18-HQ dataset~\cite{musdb}). Furthermore, we demonstrate that NeuroAMP is able to mimic conventional amplification methods and provide personalized amplification in an end-to-end manner.

To demonstrate its integration potential, we further introduce Denoising NeuroAMP, which combines NeuroAMP with a speech enhancement module, showcasing direct integration of amplification and enhancement. The development of Denoising NeuroAMP involves combining amplification and noise reduction in one module. Specifically, the model takes noisy audio input and outputs enhanced-amplified audio, and the training objective is to minimize the loss between the predicted enhanced-amplified output from the noisy input and the corresponding ground-truth clean audio. Experimental results confirm that Denoising NeuroAMP can achieve a HASPI score of 0.90 and a HASQI score of 0.59, compared to 0.89 and 0.55 for NAL-R+WDRC, and 0.85 and 0.30 for the two-stage Denoising$\to$NAL-R+WDRC baseline, which further confirms the advantages of Denoising NeuroAMP as an integrated denoising and amplification approach.

The remainder of this paper is organized as follows. First, Section \ref{sec:related} reviews related work. Second, Section \ref{sec:methodology} introduces the proposed methods. Then, Section \ref{sec:experiments} describes the experimental setup, reports the experimental results, and discusses our findings. Finally, Section \ref{sec:conclusion} concludes this work. 

\section{Related Work}
\label{sec:related}

\subsection{Prescription Fitting Formulas}
Prescription fitting formulas aim to amplify sound based on the user's patterns of hearing loss, making speech signals audible across different frequencies while maintaining comfort. These formulas can be broadly categorized into linear and non-linear approaches. Linear amplification, exemplified by methods like NAL-R and NAL-RP \cite{NAL-RP}, provides a constant gain regardless of input volume and is generally preferred for individuals with flatter hearing loss profiles, larger dynamic ranges, and less variability in dynamic range across frequencies. It is also well-suited for those with less varied lifestyles and environments \cite{Compression}. NAL-R, developed by the Australian National Acoustic Laboratories, is commonly used for mild to moderate hearing loss, while NAL-RP is tailored for severe hearing loss. In contrast, non-linear amplification, such as NAL-NL1, NAL-NL2, and DSL m[i/o], adjusts the gain dynamically based on the input volume, frequency, and the user's specific hearing loss characteristics. This makes it more suitable for individuals with sloping hearing loss, reduced dynamic ranges, significant variations in dynamic range across frequencies, and more diverse lifestyles and listening environments. NAL-NL1 aims to equalize and normalize the relative volume levels of individual frequencies while optimizing the speech intelligibility for a given volume. NAL-NL2, an improved version of NAL-NL1, considers additional factors in calculating gains, such as gender, age, and whether the native language is tonal, which may affect output volume. NAL-NL2 is also the most widely used fitting formula at present. In contrast, DSL m[i/o] is designed to prevent uncomfortable loudness while using hearing aids and to maximize the audibility of essential messages in conversations. Furthermore, these non-linear formulas utilize compression techniques to manage the reduced auditory dynamic range commonly experienced by individuals with sensorineural hearing loss.

\subsection{Hearing Aid Compression}

People with sensorineural hearing loss tend to have a reduced auditory dynamic range. For these individuals, soft sounds are inaudible, while intense sounds are perceived as loudly as they are by a normal-hearing ear. To optimize the use of the residual dynamic range and restore normal loudness perception, most modern hearing aids adopt compression. WDRC is a commonly adopted compression strategy designed to prevent sounds from being either too loud or too soft. The use of WDRC in hearing aids provides several benefits, such as ensuring consistent audibility of speech signals at safe and comfortable hearing levels \cite{WDRCbenefit1}. Moreover, WDRC involves multiple steps of audio processing, which are \cite{WDRCblock}:

\begin{itemize}
\item \textbf{Signal analysis through short-time Fourier transform (STFT):} The input audio signal undergoes signal analysis through STFT.

\item \textbf{Filterbank construction:} Frequency bins are grouped into a predefined number of filterbanks, and their volume levels are estimated.

\item \textbf{Calculation of filterbank-specific gain functions:} Gain functions specific to each filterbank are calculated based on the level estimations. 

\item \textbf{Interpolation and modification:} The gain functions are interpolated to individual frequency bins and then used to modify the STFT representation of the input signal. 

\item \textbf{Signal reconstruction:} The modified STFT representation is converted back to the time domain using the Inverse STFT.
\end{itemize}

These steps ensure that the WDRC system effectively compresses a wide range of volume levels while maintaining listener comfort. In addition, the WDRC considers the listener's hearing thresholds and dynamic range variations across frequencies, aiming to enhance speech intelligibility.

\section{Methodology}
\label{sec:methodology}

This section presents the architectures of the two proposed frameworks. 
We first introduce the NeuroAMP architecture, which outlines the core design for audio amplification using a neural network-based system. Subsequently, we introduce the Denoising NeuroAMP architecture, which is an extension of NeuroAMP that includes noise reduction capabilities.

\subsection{NeuroAMP Architecture}
\begin{figure}[!t]
\centering
\includegraphics[scale=0.57]{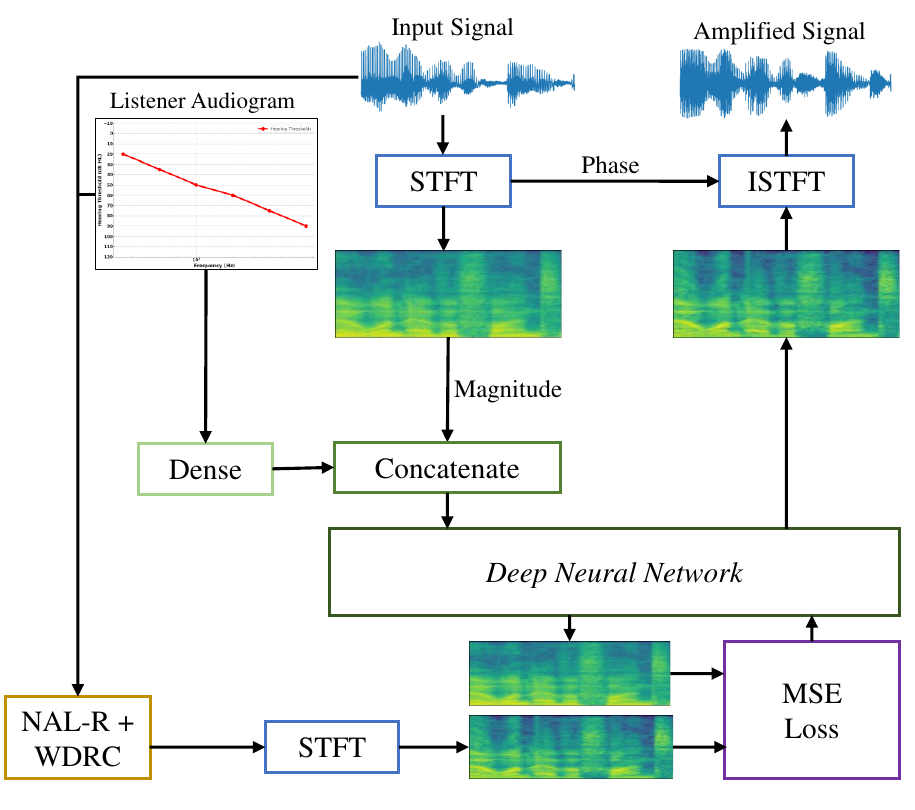}
\caption{Architecture of the NeuroAMP model, illustrating the process flow from the input signal and audiogram to feature extraction and neural network processing to the final output signal.
}
\label{fig_N}
\end{figure}

The processing workflow of NeuroAMP involves taking an audio signal and the listener's audiogram as inputs, processing them through a neural network, and producing an amplified output signal. The detailed architecture of the NeuroAMP model is illustrated in Fig.~\ref{fig_N}.
Specifically, given the input signal $y$, the STFT is applied to extract its time-frequency spectral features ($\textbf{Y} = [\textbf{y}_1, \dots, \textbf{y}_t, \dots, \textbf{y}_T]$). Simultaneously, the audiogram $z$ is transformed into a vector representation via a dense layer and then replicated into a sequence of length $T$ ($\textbf{Z} = [\textbf{z}_1, \dots, \textbf{z}_t, \dots, \textbf{z}_T]$). The two input streams are concatenated along the feature dimension and then passed to the subsequent layers of the model. In our setup, we concatenate these two sets of features in the latent space to leverage the neural network’s ability to capture semantic similarities and perform effective feature fusion. This approach has been shown to be effective for building personalized deep-learning models \cite{speaker_emb}. The feedforward process of NeuroAMP is defined as follows:

\begin{equation}
\begin{array}{c}
\mathbf{Y} = \text{STFT}(y) \\
\mathbf{Z} = \text{Dense}(z) \\
\mathbf{Concat} = [\mathbf{Y} \mid \mathbf{Z}] \\
\mathbf{\hat{Y}} = \text{NeuroAMP}(\mathbf{Concat}). \\
\end{array}
\label{eq}
\end{equation}

In this work, for the core module in NeuroAMP, we select four neural network architectures, including CNN, LSTM, CRNN, and Transformer.

\subsubsection{Convolutional Neural Network (CNN)}
Due to their ability to extract effective features from audio spectrograms~\cite{fcnse, cnn_speech}, CNNs are particularly adept at processing complex audio signals. In NeuroAMP, CNN is used to process spectral features extracted by STFT, with the goal of capturing detailed representations of the frequency components of the audio signal. The operation of CNN in the NeuroAMP model can be described by the following equations:

\begin{equation}
\begin{aligned}
\mathbf{h}_1 &= \text{ReLU}(\text{Conv}(\mathbf{Concat}, \mathbf{W}_1) + \mathbf{b}_1) \\
\mathbf{h}_2 &= \text{ReLU}(\text{Conv}(\mathbf{h}_1, \mathbf{W}_2) + \mathbf{b}_2) \\
&\vdots \\
\mathbf{h}_L &= \text{ReLU}(\text{Conv}(\mathbf{h}_{L-1}, \mathbf{W}_L) + \mathbf{b}_L) \\
\hat{\mathbf{Y}} &= \text{Dense}(\text{Flatten}(\mathbf{h}_L))
\end{aligned}
\end{equation}

where $\mathbf{W}_l$ and $\mathbf{b}_l$ are the convolutional weights and biases of the $l$-th layer, respectively, and $L$ is the number of convolutional layers. This architecture ensures a comprehensive feature extraction process, allowing NeuroAMP to characterize auditory elements more effectively.

\subsubsection{Long Short-Term Memory (LSTM)}
In the context of hearing aids, preserving the temporal structure of the audio signal is essential to maintaining speech intelligibility and naturalness. The LSTM is effective at capturing long-range dependencies and temporal patterns~\cite{lstm, lstm1, lstm2}, which is crucial for processing continuous audio signals and adapting to varying acoustic environments. The process of the LSTM in NeuroAMP is described as follows:

\begin{equation}
\begin{aligned}
\mathbf{h}_t, \mathbf{c}_t &= \text{LSTM}(\mathbf{Concat}_t, \mathbf{h}_{t-1}, \mathbf{c}_{t-1}) \\
\hat{\mathbf{Y}} &= \text{Dense}(\mathbf{h}_T),
\end{aligned}
\end{equation}
where $\mathbf{h}_t$ and $\mathbf{c}_t$ are the hidden state and cell state at time step $t$, respectively.

\subsubsection{Convolutional Recurrent Neural Network (CRNN)}
CRNN combines the strengths of CNN and RNN, making it an even better model for signal processing tasks~\cite{crnn, crnnse} that involve both spatial and temporal feature extraction. Its convolutional layers capture local spectral features, while the recurrent layers model temporal dynamics. This hybrid approach enhances the network's ability to manage audio signal complexity, thereby improving performance in tasks like speech enhancement~\cite{tan2018convolutional, hsieh2020wavecrn}. The process of the CRNN in the NeuroAMP model is described as follows:
\begin{equation}
\begin{aligned}
\mathbf{h}_c &= \text{CNN}(\mathbf{Concat}) \\
\mathbf{h}_t, \mathbf{c}_t &= \text{LSTM}(\mathbf{h}_c, \mathbf{h}_{t-1}, \mathbf{c}_{t-1}) \\
\mathbf{\hat{Y}} &= \text{Dense}(\mathbf{h}_T),
\end{aligned}
\end{equation}
where $\mathbf{h}_c$ is the output of the convolutional layer.


\subsubsection{Transformer}
The Transformer has demonstrated state-of-the-art performance in sequence modeling tasks~\cite{vaswani2017attention, transformersurvey, wolf2020transformers, baevski2020wav2vec}. Unlike RNNs, Transformer uses a self-attention mechanism~\cite{vaswani2017attention} to process entire sequences in parallel, allowing efficient handling of long-range dependencies~\cite{karitaasru}. This capability is particularly beneficial in audio signal processing~\cite{stransformer}, where capturing the global context and dependencies can enhance performance. Transformer process in the NeuroAMP model is described as follows:

\begin{equation}
\begin{aligned}
\mathbf{Q}, \mathbf{K}, \mathbf{V} &= \mathbf{W}_Q \mathbf{Concat}, \mathbf{W}_K \mathbf{Concat}, \mathbf{W}_V \mathbf{Concat} \\
\mathbf{Attention} &= \text{Softmax} \left( \frac{\mathbf{Q} \mathbf{K}^T}{\sqrt{d_k}} \right) \mathbf{V} \\
\mathbf{h} &= \text{FFN}(\mathbf{Attention}) \\
\mathbf{\hat{Y}} &= \text{Dense}(\mathbf{h}),
\end{aligned}
\end{equation}
where $\mathbf{Q}$, $\mathbf{K}$, and $\mathbf{V}$ are the query, key, and value matrices, $\mathbf{W}_Q$, $\mathbf{W}_K$, and $\mathbf{W}_V$ are the weight matrices for the query, key and value, $d_k$ is the dimension of the key, and FFN represents the feedforward network.

\subsubsection{Loss function}
To optimize NeuroAMP, we use Mean Squared Error (MSE) as the loss function. This loss function penalizes deviations between predicted ($\hat{\mathbf{Y}}$) and ground-truth ($\bar{\mathbf{Y}}$) spectral representations, guiding the network to learn spectral mappings. The loss function is calculated as:

\begin{equation}
\label{mse}
L_{\text{NeuroAMP}} = \frac{1}{T} \sum_{t=1}^T \left\| \hat{\mathbf{y}}_{t} - \bar{\mathbf{y}}_{t} \right\|_2^2,
\end{equation}

where $\bar{\mathbf{y}}_t$ and $\hat{\mathbf{y}}_t$ are the ground-truth amplified and estimated-amplified spectral representations of the $t$-th frame of the input audio signal $y$, and $T$ is the number of frames.




\subsection{Denoising NeuroAMP Architecture}

\begin{figure}[!t]
\centering
\includegraphics[scale=0.57]{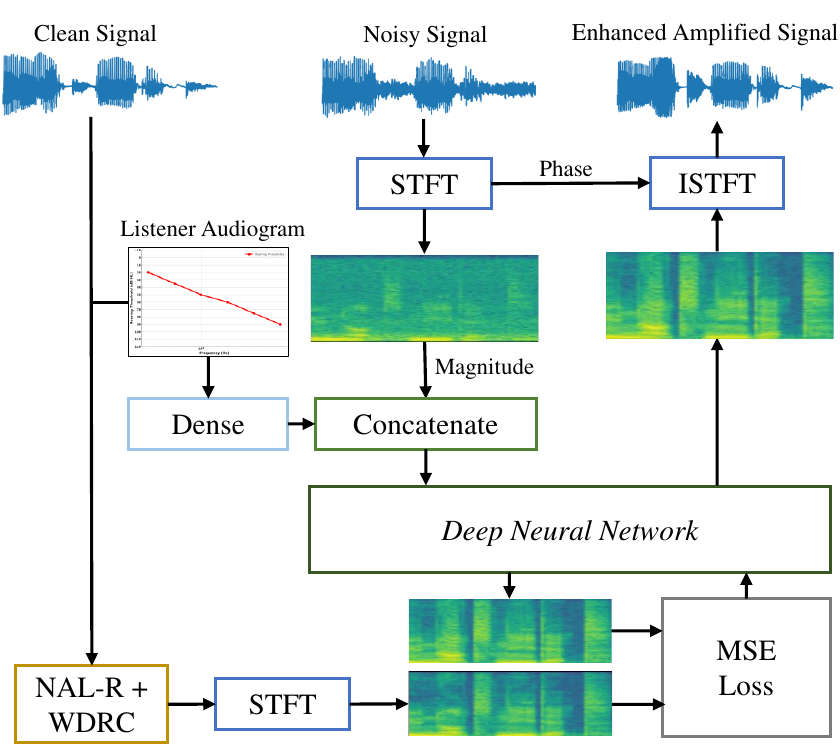}
\caption{Architecture of the Denoising NeuroAMP model, illustrating the process flow from the input signal and audiogram to feature extraction and neural network processing to the final output signal.
}
\label{fig_DN}
\end{figure}

To address the challenges of noisy environments, which severely impact the effectiveness of hearing aids, we propose the Denoising NeuroAMP system. This system is built on the NeuroAMP architecture, which allows simultaneous denoising and amplification by seamlessly integrating speech enhancement within the NeuroAMP model. The architecture of the Denoising NeuroAMP model is illustrated in Fig.~\ref{fig_DN}.

As shown, Denoising NeuroAMP takes a noisy audio signal (\textbf{X}) as input, represented as a magnitude spectrogram. Simultaneously, the listener's audiogram is processed through a dense layer to generate a vector representation corresponding to the listener's hearing profile. These two features – the noisy magnitude spectrogram and the audiogram vector – are concatenated and then fed into the core Denoising NeuroAMP neural network. This network, while conceptually similar to the original NeuroAMP model, is trained to perform speech enhancement and personalized amplification in a single unified process. Based on the noisy input characteristics and the listener’s specific hearing loss profile, Denoising NeuroAMP generates amplified audio.

The training objective of Denoising NeuroAMP, as shown in Eq. \ref{mse_den_neuro}, is to minimize the MSE between the ground-truth magnitude spectrogram (obtained by applying NAL-R and WDRC to the clean input audio) and the predicted enhanced amplified magnitude spectrogram derived from the noisy input. This ensures that the system learns to produce amplified outputs that are both enhanced and aligned with established hearing aid fitting practices.

\begin{equation}
\begin{array}{c}
\mathbf{X} = \text{STFT}(x) \\
\mathbf{Z} = \text{Dense}(z) \\
\mathbf{Concat} = [\mathbf{X} \mid \mathbf{Z}] \\
\mathbf{\hat{X}} = \text{Den-NeuroAMP}(\mathbf{Concat}). \\
\end{array}
\label{eq_concat_den_neuro}
\end{equation}

\begin{equation}
\label{mse_den_neuro}
L_{\text{Den-NeuroAMP}} = \frac{1}{T} \sum_{t=1}^T \left| \bar{\mathbf{y}}_t - \hat{\mathbf{x}}_t \right|^2,
\end{equation}

The output of the Denoising NeuroAMP network is a predicted magnitude spectrogram of the amplified enhanced signal. The phase information of the input noisy signal is combined with the predicted magnitude spectrogram to reconstruct the time-domain, enhanced-amplified audio signal through the Inverse Short-Time Fourier Transform (ISTFT). 

\section{Experiments}
\label{sec:experiments}
\subsection{Dataset}

In this work, we train the NeuroAMP model by using speech signals of two languages (English and Taiwanese Mandarin speech data) and music data. We aim to evaluate the proposed models under diverse acoustic conditions of cross-linguistic speech and music signals. The speech data contains three conditions: clean, noisy, and enhanced speech. To generate noisy speech, we added 100 types of environmental noise from~\cite{noise} to clean speech at different signal-to-noise ratio (SNR) levels. Enhanced speech was then obtained by applying two deep learning-based speech enhancement models to the noisy data. The first model is based on LSTM~\cite{lstmse}, which operates on STFT features to enhance noisy speech. The second model is based on a Fully Convolutional Network (FCN)~\cite{fcnse}, which operates on the raw waveform to enhance noisy speech. For music data, we mainly consider two categories: clean music and noisy music (i.e., music with added noise). Clean music represents the original recording, while noisy music incorporates various environmental disturbances to simulate real-world listening scenarios. 
To train the NeuroAMP model, we used TIMIT, TMHINT, and the music dataset provided by the Cadenza Challenge~\cite{cedenza1}. To test and evaluate the model's performance and generalization, we also included the unseen test set (the acoustic conditions and audio content are entirely different from the training conditions) from the VoiceBank-DEMAND and MUSDB18-HQ datasets.

\subsubsection{Training Dataset Description}
\begin{itemize}
    \item \textbf{TIMIT Dataset}: The TIMIT dataset is a well-known speech corpus used for tasks like speech recognition~\cite{timitasr, timitse} and speech enhancement~\cite{timitse, timitse1}. It includes recordings from 630 speakers representing eight major dialect regions of American English, with each speaker reading 10 phonetically rich sentences. The TIMIT training set consists of 4,620 audio files, while the test set comprises 1,690 audio files. In addition, to augment the dataset, noise signals were injected at different SNR levels (-5 dB, 0 dB, and 5 dB) into the clean audio to create noisy audio. In addition, Enhanced versions of these noisy audio samples were then generated using two SE models (LSTM and FCN). Following augmentation, a total of 13,700 training samples were prepared based on the TIMIT dataset, consisting of 3,425 clean audio, 3,425 noisy audio, and 6,850 enhanced audio.
    
    \item \textbf{TMHINT Dataset}: The TMHINT dataset was used to improve the reliability of our models by including diverse acoustic conditions. The TMHINT dataset has been widely used in various speech-related tasks, such as speech assessment~\cite{mosa} and enhancement~\cite{timitse}. Specifically, TMHINT dataset consists of recordings from multiple speakers, Each speaker provided 16 phonemically balanced sentence sets, each containing 20 sentences. Each sentence comprises ten characters representing diverse and balanced phonetic elements of Mandarin, including four primary lexical tones and a neutral tone. The total number of utterances in this dataset was 1440, which we divided into training and testing sets using an 80/20 ratio, ensuring that speakers in the test set were not used in training. In addition, we expanded the TMHINT dataset by adding environmental noises at different SNR levels (-5 dB, 0 dB, and 5 dB). We also generated enhanced speech from the noisy speech using LSTM and FCN SE models. We then generated reference data using NAL-R+WDRC, and split this equally between the training and testing sets.
    
    \item \textbf{MUSIC Dataset}: The music dataset from the Cadenza Challenge~\cite{cedenza1,cedenza2} was used to train the NeuroAMP model. To simulate real-world listening conditions, environmental noise was added at SNR levels of $-5$, $0$, and $5$~dB. This augmentation mimics scenarios where music is played with background noise. In total, we used 11{,}856 training samples: 5{,}928 clean and 5{,}928 noisy.


\subsubsection{Testing Dataset Description}
To evaluate the models' performance, we first utilized the test sets from the same datasets used during training: TIMIT, TMHINT, and the Cadenza Challenge music data. Our evaluation methodology involved testing the models on these datasets with added unseen noises at varying SNR levels of -6 dB, 0 dB, and 6 dB. This approach allows us to assess how well the models handle different noise conditions. 

In addition, we included separate speech and music test sets that were not part of the training data (VoiceBank-DEMAND, and MUSDB18-HQ) datasets to evaluate the models' ability to generalize to the unseen environments. The detailed information for the two unseen datasets is as follows:

    \item \textbf{VoiceBank-DEMAND Dataset}: This dataset has been widely employed in various speech-related tasks, such as speech enhancement~\cite{matricse, vctkse} and recognition~\cite{vctkasr}. The dataset contains recordings of English speech utterances spoken by multiple speakers, capturing diverse accents and conditions. The test set includes both clean and noisy samples, each containing 824 utterances from 2 speakers. The noisy samples cover four SNR levels: 17.5 dB, 12.5 dB, 7.5 dB, and 2.5 dB, representing different degrees of noise added to the speech signals.

    \item \textbf{MUSDB18-HQ Dataset}~\cite{musdb}: The evaluation set was expanded by including the MUSDB18 dataset, another prominent resource in the field of audio processing~\cite{musdb1, musdb2}. The MUSDB18 dataset comprises multi-track music recordings, facilitating source separation and music transcription tasks. By incorporating the MUSDB18 dataset into our evaluation process, we aim to assess the performance of our models in handling music signals.

\end{itemize}

\subsubsection{Denoising NeuroAMP Dataset}
We train and evaluate Denoising NeuroAMP on the VoiceBank-DEMAND dataset, which combines clean speech from VoiceBank with diverse environmental noises from DEMAND at various SNRs. Using the official split (11,572 training utterances from 28 speakers; 824 test utterances from 2 speakers), this setup provides clean/noisy pairs ideal for jointly learning speech denoising and personalized amplification.

\subsubsection{Hearing Loss Patterns}
In our study, we used audiograms from the Clarity Challenge~\cite{akeroyd20232nd}, which provides comprehensive samples for both training and testing. These audiograms are essential for evaluating hearing abilities across a standard set of frequencies. We specifically examine audiograms that cover six fundamental frequencies: 250, 500, 1000, 2000, 4000, and 6000 Hz. These frequencies are important for assessing speech intelligibility and understanding the auditory challenges faced by individuals with hearing loss. We follow the approach described by Alshuaib et al.~\cite{hlp} to generate hearing loss patterns, which offers a systematic method to categorize audiometric data into distinct patterns.

\subsection{Experimental Setup}

This section details the experimental setups for both the NeuroAMP and Denoising NeuroAMP models.

\subsubsection{NeuroAMP Setup}

In the NeuroAMP experimental setup, the input signal is preprocessed with a 512-point STFT, a Hamming window of 32 ms, and a hop of 16 ms. The audiogram input, represented as a 1x6 matrix of hearing thresholds at specific frequencies, is processed through dense layers to create a refined representation. This processed audiogram information is then concatenated with the spectral features to form the input vector for further modeling.

The various model architectures and their configurations employed in our study are as follows:\footnote{https://github.com/Shafique-Khattak/NeuroAMP}

\begin{itemize}
\item \textbf{CNN:} The CNN configuration in NeuroAMP includes four convolutional layers with filter sizes of 32, 64, 128, and 256. This setup aims to extract detailed spatial features from the audio signals.
\item \textbf{LSTM:} The LSTM model configuration includes two layers, each with 256 units, designed to capture temporal dependencies in the audio signal.
\item \textbf{CRNN:} The CRNN model combines four convolutional layers with 16, 32, 64, and 128 filter sizes, followed by two LSTM layers, each with 256 units. This structure is particularly effective for handling both spectral features and temporal sequences.
\item \textbf{Transformer:} The Transformer model includes four encoder blocks, each with 16 heads. These blocks use an attention mechanism that helps the model better understand subtle details in complex audio settings.
\end{itemize}

Each model concludes with a 257-unit dense layer, corresponding to the number of positive frequency bins in a 512-point STFT ($512/2 + 1 = 257$). This layer directly predicts the per-bin amplified spectrum for every frame. Because the audiogram embedding is concatenated with the STFT features before entering the neural network, the output is inherently conditioned on the listener’s specific hearing profile.
The models were trained for $100$ epochs using the Adam optimizer with a learning rate of $1\times10^{-4}$, $\beta_1 = 0.9$, $\beta_2 = 0.999$, and $\epsilon = 1\times10^{-7}$. Training was performed with a batch size of $1$, and early stopping on the validation loss was applied to prevent overfitting.

\subsubsection{Denoising NeuroAMP Setup}

For the Denoising NeuroAMP experiments, the setup largely mirrors that of NeuroAMP, with a few key distinctions. Similarly, the input signals are preprocessed using a 512-point STFT, a 32 ms Hamming window, and a 16 ms hop size. The listener's audiogram, represented as a 1x6 vector, is processed through a dense layer and concatenated with the spectral features of the noisy input signal.
In the Denoising NeuroAMP, we specifically employ an LSTM-based neural network architecture. The LSTM model includes two layers, each with 256 units. This configuration is chosen for its effectiveness in capturing temporal dependencies in the audio signal, which is crucial for both speech enhancement and personalized amplification.
The training target for Denoising NeuroAMP is generated by applying the NAL-R prescription formula combined with WDRC to the clean version of the input audio signal. We use the MSE as the loss function, calculated between the predicted magnitude spectrogram and the target magnitude spectrogram.

The Denoising NeuroAMP model is optimized using the same Adam optimizer configuration as NeuroAMP (learning rate of $1\times10^{-4}$, $\beta_1 = 0.9$, $\beta_2 = 0.999$, and $\epsilon = 1\times10^{-7}$). Training is performed with a batch size of $1$ for $100$ epochs, with an early stop on validation loss to prevent overfitting.

\subsection{Evaluation Results}

This section outlines the evaluation strategy, focusing on assessing the intelligibility and quality of amplified speech processed by the proposed NeuroAMP model and NAL-R+WDRC (baseline). We employ two established metrics in hearing aid auditory research: HASPI and HASQI for speech, as well as the HAAQI for music. These metrics are important for quantifying the model's impact on both speech and music quality. For benchmarking purposes, we compare the metrics derived from NeuroAMP with those obtained from the NAL-R+WDRC processed signal, using three statistical measures: Linear Correlation Coefficient (LCC), SRCC, and MSE. Higher LCC values, higher SRCC values, and lower MSE values indicate greater similarity between NeuroAMP and NAL-R+WDRC.

To provide a robust assessment of model performance, all metrics are reported as mean ± standard deviation. Statistical significance between models was determined using paired t-tests, with p-values below 0.05 considered significant. Significance is denoted in the results tables using appropriate symbols. Furthermore, we explore the relationship between model parameter scale and performance through two experimental settings: a basic parameter configuration for all neural networks to establish a baseline (denoted as basic (same)) and an optimized configuration to evaluate the full potential of each architecture (denoted as optimized).

\subsubsection{Impact of Model Parameter Scale on Performance}

\textbf{Table I} summarizes the number of parameters for each model under basic (same) and optimized settings. This comparison aims to evaluate the trade-offs between model complexity and performance.

\begin{table}[t]
  \centering
  \caption{Number of Parameters of Basic and Optimized for Each Model}
  \label{tab:combined_parameters}
  \resizebox{\columnwidth}{!}{ 
  \begin{tabular}{lcc}
    \toprule
    Model & Parameters (Basic(Same)) & Parameters (Optimized) \\
    \midrule
    CNN & 498,411 & 672,235 \\
    LSTM & 498,224 & 1,134,193 \\
    CRNN & 496,408 & 1,503,595 \\
    Transformer & 499,884 & 2,456,398 \\
    \bottomrule
  \end{tabular}
  }
\end{table}

\textbf{Table II} presents the results under the basic (same) parameter setting, reported as mean ± standard deviation with statistically significant differences indicated. Under these constraints, the Transformer model consistently exhibited superior performance across most datasets and metrics. It achieved the highest LCC and SRCC values and the lowest MSE, indicating its inherent architectural efficiency and ability to generalize well even with limited parameters.

\begin{table*}[t]
  \centering
  \caption{LCC, SRCC, and MSE Results of Different Models on Speech Datasets (mean $\pm$ std, basic (same) Parameter Setting).}
  \label{tab:performance1}
  \resizebox{\textwidth}{!}{
  \begin{tabular}{lccccccccccc}
    \toprule
    \multirow{2}{*}{Metric} & \multirow{2}{*}{Model} & \multicolumn{3}{c}{TIMIT} & \multicolumn{3}{c}{TMHINT} & \multicolumn{3}{c}{VoiceBank-DEMAND (Unseen)} \\
    \cmidrule(lr){3-5}\cmidrule(lr){6-8}\cmidrule(lr){9-11}
    & & LCC $\uparrow$ & SRCC $\uparrow$ & MSE $\downarrow$ & LCC $\uparrow$ & SRCC $\uparrow$ & MSE $\downarrow$ & LCC $\uparrow$ & SRCC $\uparrow$ & MSE $\downarrow$ \\
    \midrule
    \multirow{4}{*}{HASPI} 
    & CNN         & 0.971 $\pm$ 0.006 & 0.984 $\pm$ 0.003 & 0.006 $\pm$ 0.001 & 0.986 $\pm$ 0.003 & 0.982 $\pm$ 0.004 & 0.002 $\pm$ 0.001 & 0.931 $\pm$ 0.018 & 0.970 $\pm$ 0.002 & 0.004 $\pm$ 0.001 \\
    & LSTM        & 0.982 $\pm$ 0.004 & 0.989 $\pm$ 0.002 & 0.004 $\pm$ 0.001 & 0.988 $\pm$ 0.004 & 0.984 $\pm$ 0.003 & 0.002 $\pm$ 0.000 & 0.944 $\pm$ 0.015 & 0.971 $\pm$ 0.002 & 0.003 $\pm$ 0.001 \\   
    & CRNN        & 0.973 $\pm$ 0.003 & 0.978 $\pm$ 0.002 & 0.006 $\pm$ 0.001 & 0.978 $\pm$ 0.008 & 0.972 $\pm$ 0.004 & 0.004 $\pm$ 0.001 & 0.921 $\pm$ 0.008 & 0.943 $\pm$ 0.003 & 0.006 $\pm$ 0.001 \\
    & Transformer & 0.981 $\pm$ 0.006$^{*\ddagger}$ & 0.990 $\pm$ 0.002$^{*\ddagger}$ & 0.004 $\pm$ 0.001$^{*\ddagger}$ 
                   & 0.990 $\pm$ 0.002$^{*\dagger\ddagger}$ & 0.989 $\pm$ 0.003$^{*\dagger\ddagger}$ & 0.001 $\pm$ 0.000$^{*\dagger\ddagger}$ 
                   & 0.954 $\pm$ 0.019$^{*\dagger\ddagger}$ & 0.981 $\pm$ 0.002$^{*\dagger\ddagger}$ & 0.002 $\pm$ 0.001$^{*\dagger\ddagger}$ \\
    \midrule
    \multirow{4}{*}{HASQI} 
    & CNN         & 0.994 $\pm$ 0.001 & 0.976 $\pm$ 0.006 & 0.001 $\pm$ 0.000 & 0.995 $\pm$ 0.001 & 0.976 $\pm$ 0.006 & 0.001 $\pm$ 0.000 & 0.989 $\pm$ 0.001 & 0.982 $\pm$ 0.003 & 0.002 $\pm$ 0.000 \\
    & LSTM        & 0.997 $\pm$ 0.000 & 0.986 $\pm$ 0.002 & 0.001 $\pm$ 0.000 & 0.995 $\pm$ 0.001 & 0.977 $\pm$ 0.007 & 0.001 $\pm$ 0.000 & 0.995 $\pm$ 0.000 & 0.992 $\pm$ 0.001 & 0.002 $\pm$ 0.000 \\
    & CRNN        & 0.990 $\pm$ 0.001 & 0.975 $\pm$ 0.002 & 0.008 $\pm$ 0.001 & 0.988 $\pm$ 0.001 & 0.965 $\pm$ 0.008 & 0.011 $\pm$ 0.001 & 0.985 $\pm$ 0.001 & 0.978 $\pm$ 0.002 & 0.015 $\pm$ 0.001 \\
    & Transformer & 0.997 $\pm$ 0.001$^{*\ddagger}$ & 0.985 $\pm$ 0.003$^{*\ddagger}$ & 0.001 $\pm$ 0.000$^{*\ddagger}$ 
                   & 0.996 $\pm$ 0.001$^{*\ddagger}$ & 0.979 $\pm$ 0.006$^{*\ddagger}$ & 0.001 $\pm$ 0.000$^{*\ddagger}$ 
                   & 0.992 $\pm$ 0.001$^{*\dagger\ddagger}$ & 0.987 $\pm$ 0.002$^{*\dagger\ddagger}$ & 0.001 $\pm$ 0.000$^{*\dagger\ddagger}$ \\
    \bottomrule
  \end{tabular}
  }
  \vspace{1mm}
  {\raggedright
  Significance symbols for the Transformer: $^{*}$ vs. CNN ($p<0.05$), $^{\dagger}$ vs. LSTM ($p<0.05$), $^{\ddagger}$ vs. CRNN ($p<0.05$).}
  \par
\end{table*}

\textbf{Table III} shows the results after optimizing the parameters, also reported as mean ± standard deviation and including significance testing. With increased parameter counts, almost all models exhibited consistent performance improvements. Notably, the CNN and LSTM models showed substantial gains in their ability to generalize, particularly on the TMHINT dataset. The CRNN model, which initially struggled under the minimal setting, demonstrated a notable improvement and achieved comparable performance after tuning. Furthermore, the Transformer model achieves the best performance and further improves out-of-domain generalization in the unseen VoiceBank-DEMAND dataset, with several improvements being statistically significant.

\begin{table*}[t]
  \centering
  \caption{LCC, SRCC, and MSE Results of Different Models on Speech Datasets (mean $\pm$ std, Optimized Parameter Setting).}
  \label{tab:performance}
  \resizebox{\textwidth}{!}{
  \begin{tabular}{lccccccccccc}
    \toprule
    \multirow{2}{*}{Metric} & \multirow{2}{*}{Model} & \multicolumn{3}{c}{TIMIT} & \multicolumn{3}{c}{TMHINT} & \multicolumn{3}{c}{VoiceBank-DEMAND (Unseen)} \\
    \cmidrule(lr){3-5}\cmidrule(lr){6-8}\cmidrule(lr){9-11}
    & & LCC $\uparrow$ & SRCC $\uparrow$ & MSE $\downarrow$ & LCC $\uparrow$ & SRCC $\uparrow$ & MSE $\downarrow$ & LCC $\uparrow$ & SRCC $\uparrow$ & MSE $\downarrow$ \\
    \midrule
    \multirow{4}{*}{HASPI} 
    & CNN         & 0.968 $\pm$ 0.006 & 0.982 $\pm$ 0.003 & 0.006 $\pm$ 0.001 & 0.982 $\pm$ 0.004 & 0.979 $\pm$ 0.004 & 0.003 $\pm$ 0.001 & 0.920 $\pm$ 0.022 & 0.962 $\pm$ 0.003 & 0.005 $\pm$ 0.001 \\
    & LSTM        & 0.982 $\pm$ 0.003 & 0.989 $\pm$ 0.002 & 0.004 $\pm$ 0.001 & 0.987 $\pm$ 0.005 & 0.982 $\pm$ 0.003 & 0.002 $\pm$ 0.000 & 0.943 $\pm$ 0.013 & 0.971 $\pm$ 0.002 & 0.003 $\pm$ 0.001 \\   
    & CRNN        & 0.986 $\pm$ 0.003 & 0.992 $\pm$ 0.001 & 0.003 $\pm$ 0.001 & 0.989 $\pm$ 0.003 & 0.984 $\pm$ 0.002 & 0.002 $\pm$ 0.000 & 0.951 $\pm$ 0.015 & 0.978 $\pm$ 0.002 & 0.003 $\pm$ 0.001 \\
    & Transformer & 0.987 $\pm$ 0.004$^{*\dagger}$ & 0.992 $\pm$ 0.001$^{*\dagger}$ & 0.003 $\pm$ 0.001$^{*\dagger}$ & 0.992 $\pm$ 0.004$^{*\dagger\ddagger}$ & 0.991 $\pm$ 0.003$^{*\dagger\ddagger}$ & 0.001 $\pm$ 0.000$^{*\dagger\ddagger}$ & 0.948 $\pm$ 0.018$^{*}$ & 0.981 $\pm$ 0.002$^{*\dagger\ddagger}$ & 0.003 $\pm$ 0.001$^{*\dagger}$ \\
    \midrule
    \multirow{4}{*}{HASQI} 
    & CNN         & 0.994 $\pm$ 0.001 & 0.976 $\pm$ 0.006 & 0.002 $\pm$ 0.000 & 0.993 $\pm$ 0.001 & 0.974 $\pm$ 0.005 & 0.002 $\pm$ 0.000 & 0.987 $\pm$ 0.001 & 0.977 $\pm$ 0.002 & 0.004 $\pm$ 0.000 \\
    & LSTM        & 0.996 $\pm$ 0.001 & 0.984 $\pm$ 0.003 & 0.002 $\pm$ 0.000 & 0.996 $\pm$ 0.001 & 0.979 $\pm$ 0.006 & 0.002 $\pm$ 0.000 & 0.994 $\pm$ 0.000 & 0.991 $\pm$ 0.001 & 0.003 $\pm$ 0.000 \\
    & CRNN        & 0.997 $\pm$ 0.000 & 0.986 $\pm$ 0.003 & 0.001 $\pm$ 0.000 & 0.995 $\pm$ 0.002 & 0.978 $\pm$ 0.007 & 0.001 $\pm$ 0.000 & 0.995 $\pm$ 0.001 & 0.992 $\pm$ 0.001 & 0.002 $\pm$ 0.000 \\
    & Transformer & 0.998 $\pm$ 0.000$^{*\dagger\ddagger}$ & 0.990 $\pm$ 0.001$^{*\dagger\ddagger}$ & 0.000 $\pm$ 0.000$^{*\dagger\ddagger}$ & 0.997 $\pm$ 0.002$^{*\dagger\ddagger}$ & 0.981 $\pm$ 0.005$^{*\ddagger}$ & 0.001 $\pm$ 0.000$^{*\dagger\ddagger}$ & 0.994 $\pm$ 0.001$^{*}$ & 0.991 $\pm$ 0.001$^{*}$ & 0.001 $\pm$ 0.000$^{*\dagger\ddagger}$ \\
    \bottomrule
  \end{tabular}
  }
  
  \vspace{1mm}
  {\raggedright
Significance symbols for the Transformer: $^{*}$ vs. CNN ($p<0.05$), $^{\dagger}$ vs. LSTM ($p<0.05$), $^{\ddagger}$ vs. CRNN ($p<0.05$).
  \par}
\end{table*}

\textbf{Table IV} provides a detailed performance evaluation of CNN, LSTM, CRNN, and Transformer models on the VoiceBank-DEMAND dataset (unseen) under different speech conditions: clean, noisy, and enhanced (using FCN and LSTM). The CRNN model excelled in clean and enhanced conditions, achieving high LCC and SRCC scores along with notably low MSE values. The Transformer model also performs well, particularly in clean and enhanced conditions.

\begin{table*}[t]
\centering
\caption{Detailed evaluation metrics for clean, noisy, and enhanced speech conditions (mean $\pm$ std)}
\label{tab:detail_performance_combined}
\resizebox{\textwidth}{!}{
\begin{tabular}{lclcccccccc}
\toprule
\multirow{4}{*}{Metric} & \multirow{4}{*}{Model} & \multicolumn{3}{c}{Clean} & \multicolumn{3}{c}{Enhanced} & \multicolumn{3}{c}{Noisy} \\
\cmidrule(lr){3-5}\cmidrule(lr){6-8}\cmidrule(lr){9-11}
& & LCC $\uparrow$ & SRCC $\uparrow$ & MSE $\downarrow$ & LCC $\uparrow$ & SRCC $\uparrow$ & MSE $\downarrow$ & LCC $\uparrow$ & SRCC $\uparrow$ & MSE $\downarrow$ \\
\midrule
\multirow{4}{*}{HASPI}
& CNN         & 0.932 $\pm$ 0.032 & 0.963 $\pm$ 0.007 & 0.001 $\pm$ 0.001 & 0.893 $\pm$ 0.028 & 0.917 $\pm$ 0.011 & 0.007 $\pm$ 0.002 & 0.961 $\pm$ 0.013 & 0.977 $\pm$ 0.004 & 0.003 $\pm$ 0.001 \\
& LSTM        & 0.950 $\pm$ 0.018 & 0.974 $\pm$ 0.006 & 0.001 $\pm$ 0.000 & 0.923 $\pm$ 0.024 & 0.942 $\pm$ 0.008 & 0.005 $\pm$ 0.002 & 0.976 $\pm$ 0.007 & 0.982 $\pm$ 0.005 & 0.002 $\pm$ 0.001 \\
& CRNN        & 0.956 $\pm$ 0.025 & 0.981 $\pm$ 0.004 & 0.001 $\pm$ 0.000 & 0.933 $\pm$ 0.019 & 0.951 $\pm$ 0.006 & 0.004 $\pm$ 0.002 & 0.982 $\pm$ 0.003 & 0.988 $\pm$ 0.002 & 0.001 $\pm$ 0.000 \\
& Transformer & 0.951 $\pm$ 0.025 & 0.986 $\pm$ 0.003$^{*}$ & 0.001 $\pm$ 0.001 & 0.931 $\pm$ 0.027 & 0.958 $\pm$ 0.009$^{*\dagger}$ & 0.004 $\pm$ 0.002 & 0.979 $\pm$ 0.010$^{*}$ & 0.987 $\pm$ 0.004$^{*}$ & 0.001 $\pm$ 0.001$^{*}$ \\
\midrule
\multirow{4}{*}{HASQI}
& CNN         & 0.749 $\pm$ 0.067 & 0.380 $\pm$ 0.073 & 0.009 $\pm$ 0.001 & 0.957 $\pm$ 0.004 & 0.954 $\pm$ 0.006 & 0.002 $\pm$ 0.000 & 0.987 $\pm$ 0.002 & 0.985 $\pm$ 0.003 & 0.002 $\pm$ 0.000 \\
& LSTM        & 0.912 $\pm$ 0.023 & 0.795 $\pm$ 0.043 & 0.007 $\pm$ 0.001 & 0.980 $\pm$ 0.002 & 0.979 $\pm$ 0.002 & 0.001 $\pm$ 0.000 & 0.994 $\pm$ 0.001 & 0.993 $\pm$ 0.001 & 0.001 $\pm$ 0.000 \\
& CRNN        & 0.922 $\pm$ 0.017 & 0.838 $\pm$ 0.019 & 0.006 $\pm$ 0.000 & 0.979 $\pm$ 0.003 & 0.979 $\pm$ 0.003 & 0.001 $\pm$ 0.000 & 0.994 $\pm$ 0.001 & 0.993 $\pm$ 0.001 & 0.001 $\pm$ 0.000 \\
& Transformer & 0.896 $\pm$ 0.023$^{*\dagger\ddagger}$ & 0.802 $\pm$ 0.025$^{*\ddagger}$ & 0.002 $\pm$ 0.000$^{*\dagger\ddagger}$ & 0.981 $\pm$ 0.003$^{*}$ & 0.981 $\pm$ 0.003$^{*}$ & 0.001 $\pm$ 0.000$^{*\dagger\ddagger}$ & 0.993 $\pm$ 0.002$^{*}$ & 0.991 $\pm$ 0.002$^{*}$ & 0.001 $\pm$ 0.000$^{*\dagger\ddagger}$ \\
\bottomrule
\end{tabular}
}
\vspace{1mm}
\begin{minipage}{\textwidth}
\raggedright
Significance symbols for the Transformer: $^{*}$ vs. CNN, $^{\dagger}$ vs. LSTM, $^{\ddagger}$ vs. CRNN $(p < 0.05)$. Enhanced results are averaged over both enhancement models.
\end{minipage}
\end{table*}

\subsubsection{Hyperparameter Sensitivity Analysis}

To further evaluate the stability and generalizability of the proposed method, we conducted two complementary hyperparameter sensitivity analyses—one focusing on model scale (Small, Medium, Large) and one on model depth (1–4 Transformer blocks). For the scale experiment, variants differ in number of layers and hidden units; for the depth experiment, we fixed other parameters and varied only the block count. All models were evaluated on the unseen VoiceBank-DEMAND dataset using the HASPI and HASQI metrics. We report Pearson’s LCC, Spearman’s SRCC, and MSE, each with 95 \% confidence intervals.

\begin{figure}[!t]
\centering
\includegraphics[scale=0.21]{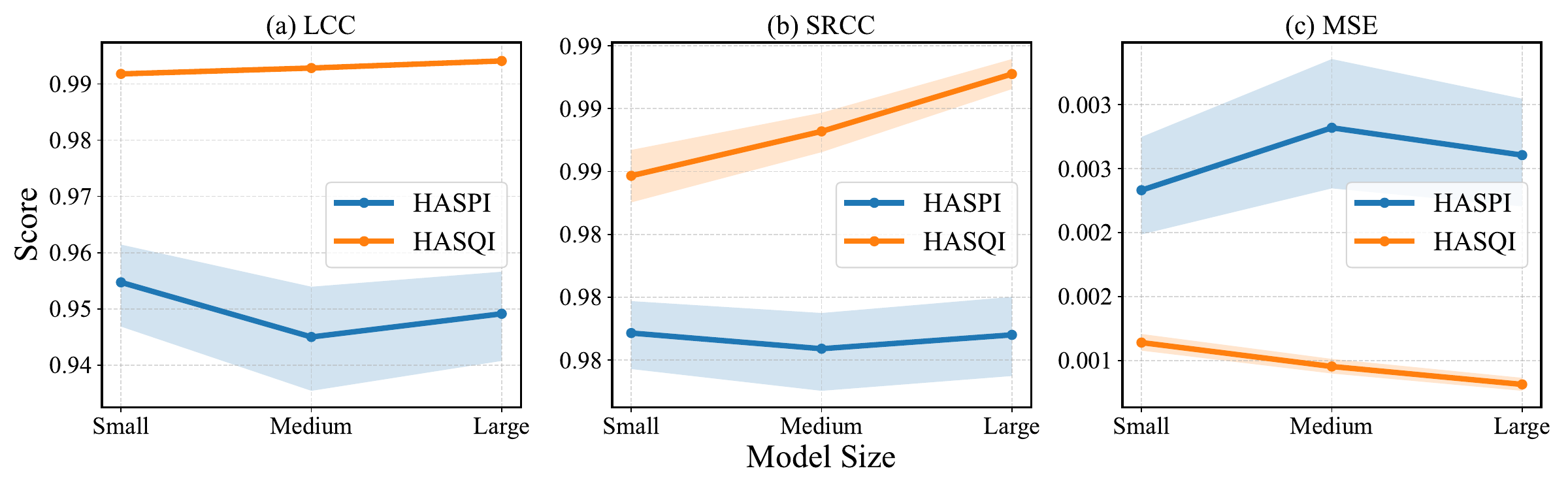}
\caption{Performance of the Transformer models of varying sizes (Small/Medium/Large) on HASPI and HASQI. (a) LCC, (b) SRCC, and (c) MSE. Solid lines denote the mean; shaded bands indicate $95\%$ confidence intervals.}
\label{hyperparameter_size}
\end{figure}

\begin{figure}[!t]
\centering
\includegraphics[scale=0.21]{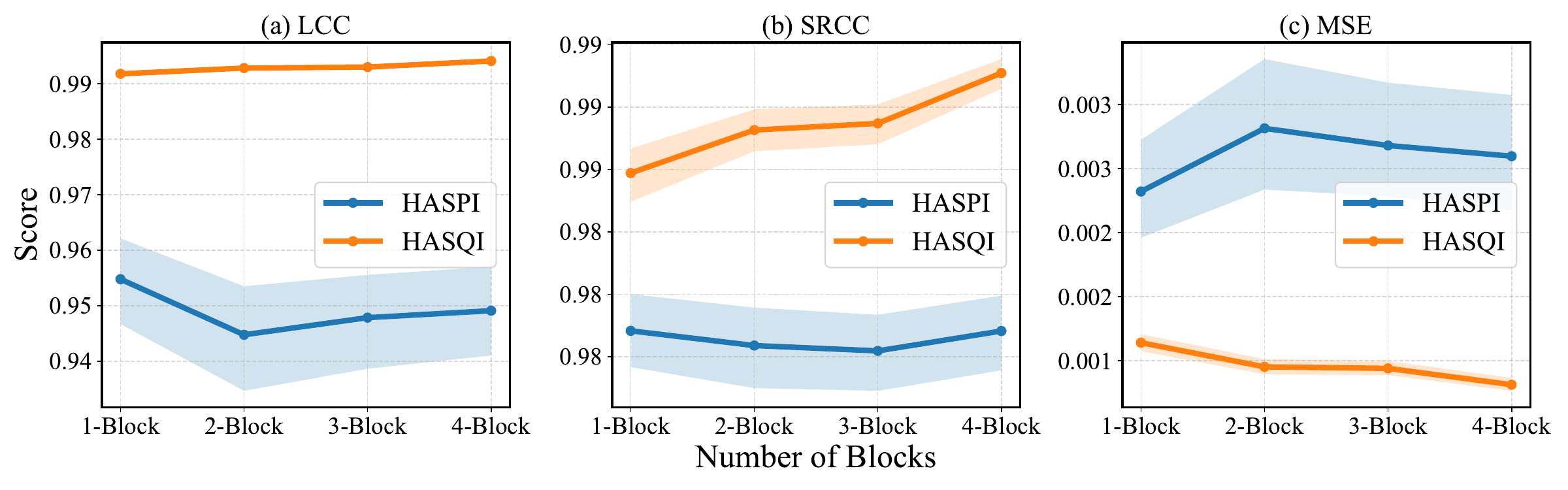}
\caption{Performance of the Transformer models with varying number of blocks (1–4) on HASPI and HASQI. (a) LCC, (b) SRCC, and (c) MSE. Solid lines denote the mean; shaded bands indicate 95\% confidence intervals.}
\label{hyperparameter_block}
\end{figure}

Figures~\ref{hyperparameter_size} (scale) and~\ref{hyperparameter_block} (depth) summarize these results. Across both analyses and all metrics, performance remains remarkably stable, with only marginal gains observed when increasing the model size beyond Medium or adding more than 2–3 blocks. The nearly overlapping confidence intervals indicate that our Transformer-based system is robust to reasonable variations in both size and depth, supporting its practical deployability and generalization capability.

\begin{figure}[!t]
\centering
\includegraphics[scale=0.29]{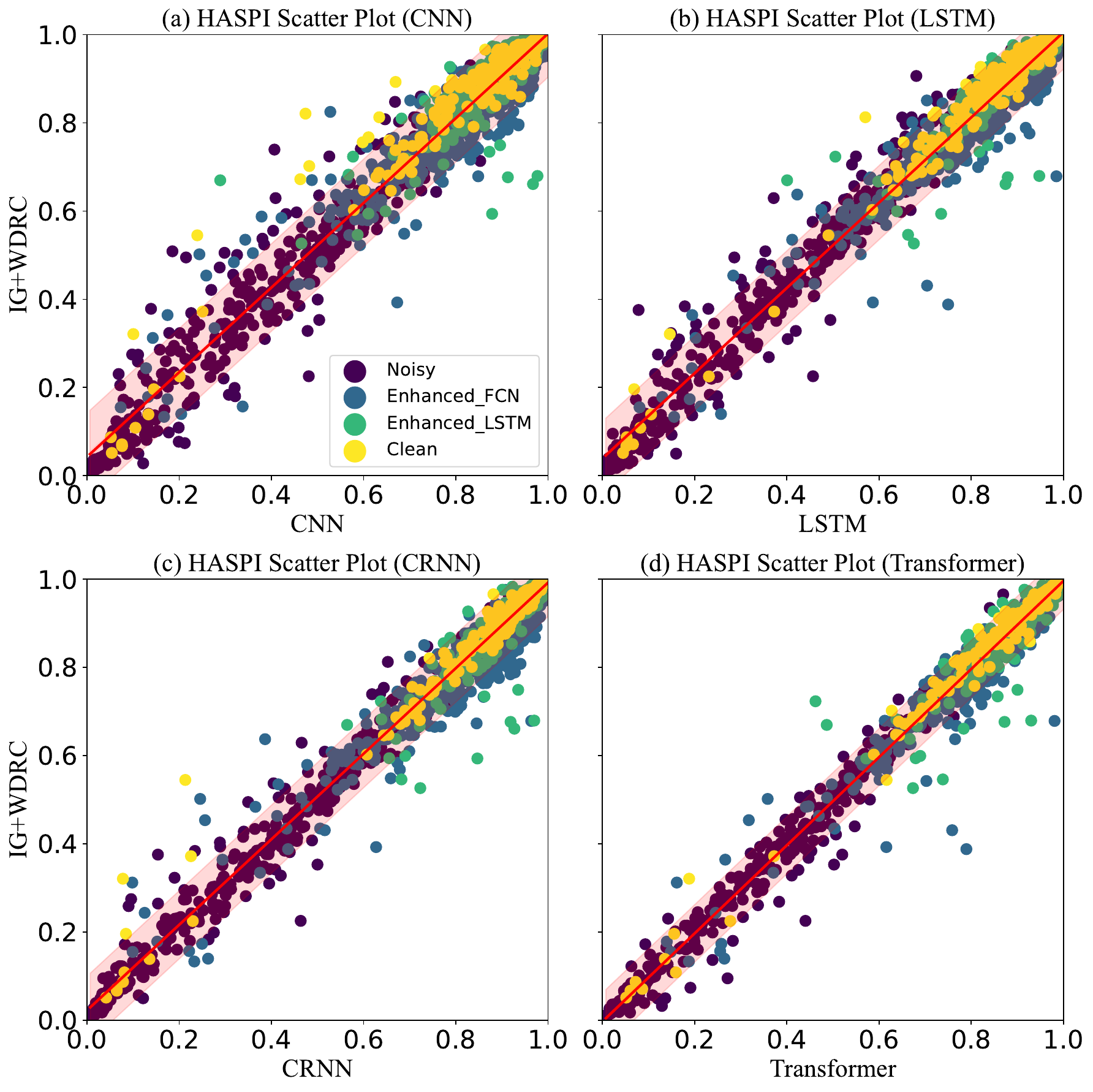}
\caption{TMHINT dataset scatter plots for HASPI scores (a) CNN model architecture (b) LSTM model architecture (c) CRNN model architecture (d) Transformer model architecture}
\label{fig_6}
\end{figure}

\begin{figure}[!t]
\centering
\includegraphics[scale=0.29]{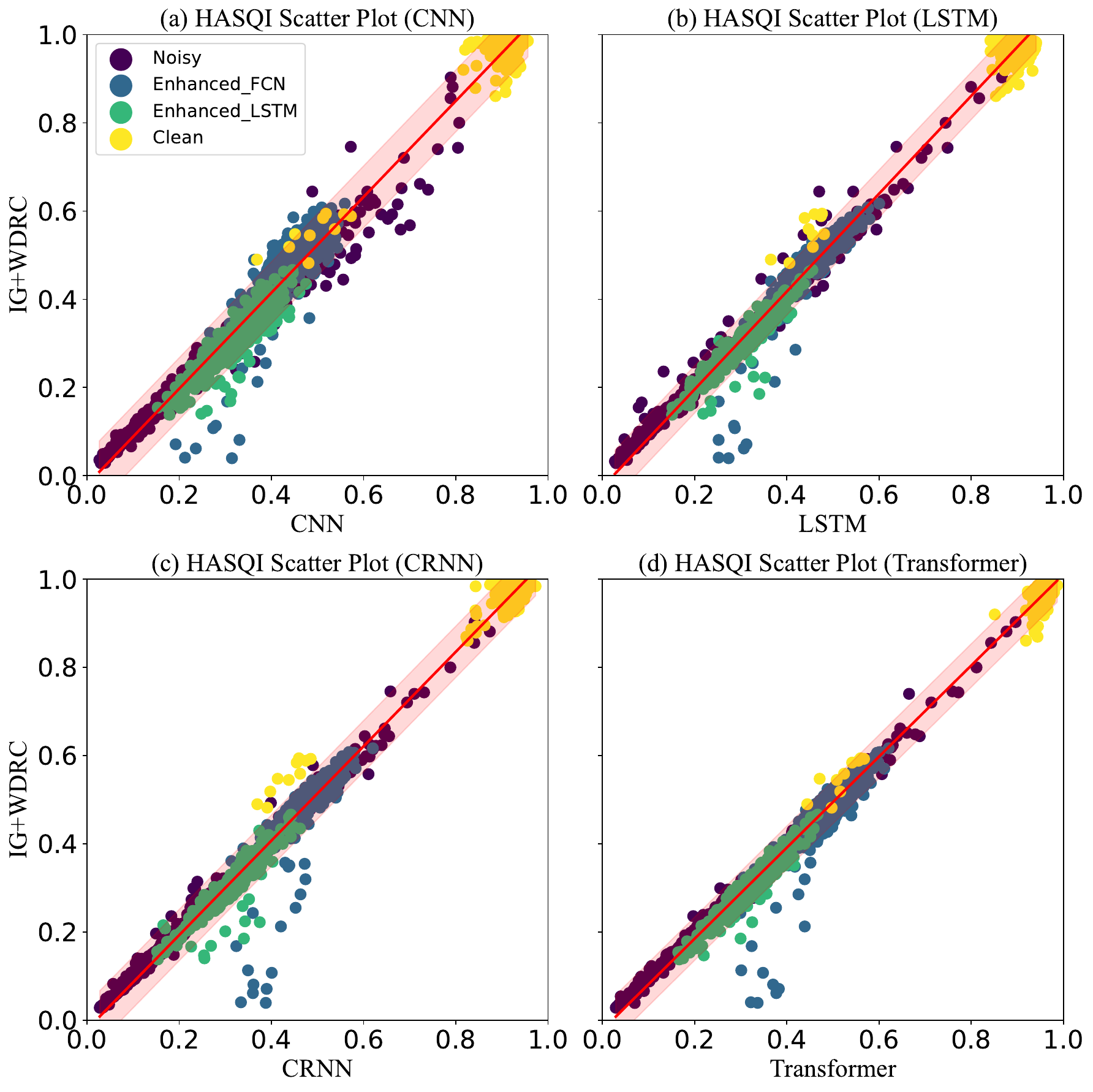}
\caption{TMHINT dataset scatter plots for HASQI scores (a) CNN model architecture (b) LSTM model architecture (c) CRNN model architecture (d) Transformer model architecture}
\label{fig_7}
\end{figure}

\begin{figure}[!t]
\centering
\includegraphics[scale=0.29]{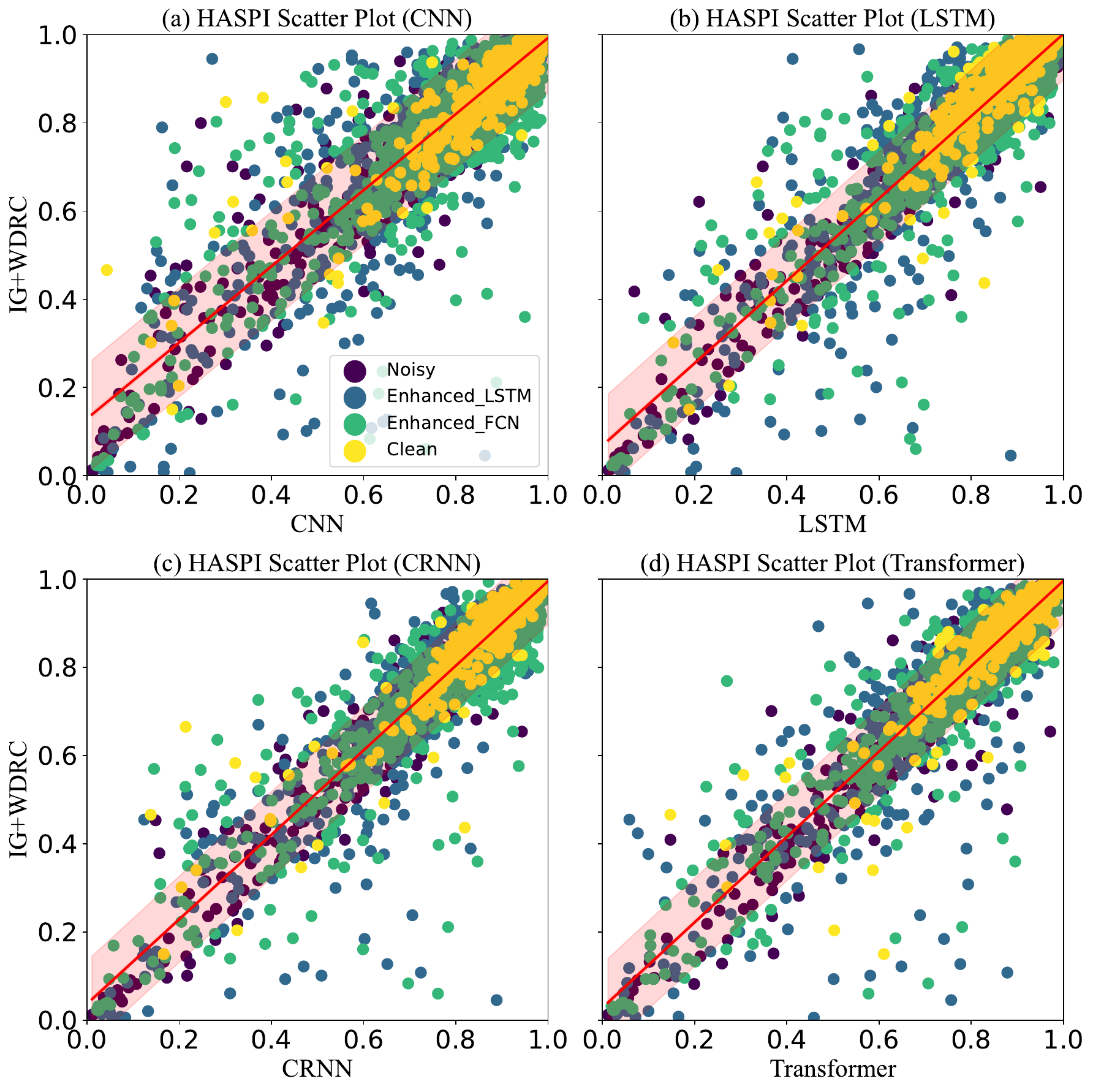}
\caption{VoiceBank-DEMAND dataset scatter plots for HASPI scores (a) CNN model architecture (b) LSTM model architecture (c) CRNN model architecture (d) Transformer model architecture}
\label{fig_8}
\end{figure}

\begin{figure}[!t]
\centering
\includegraphics[scale=0.29]{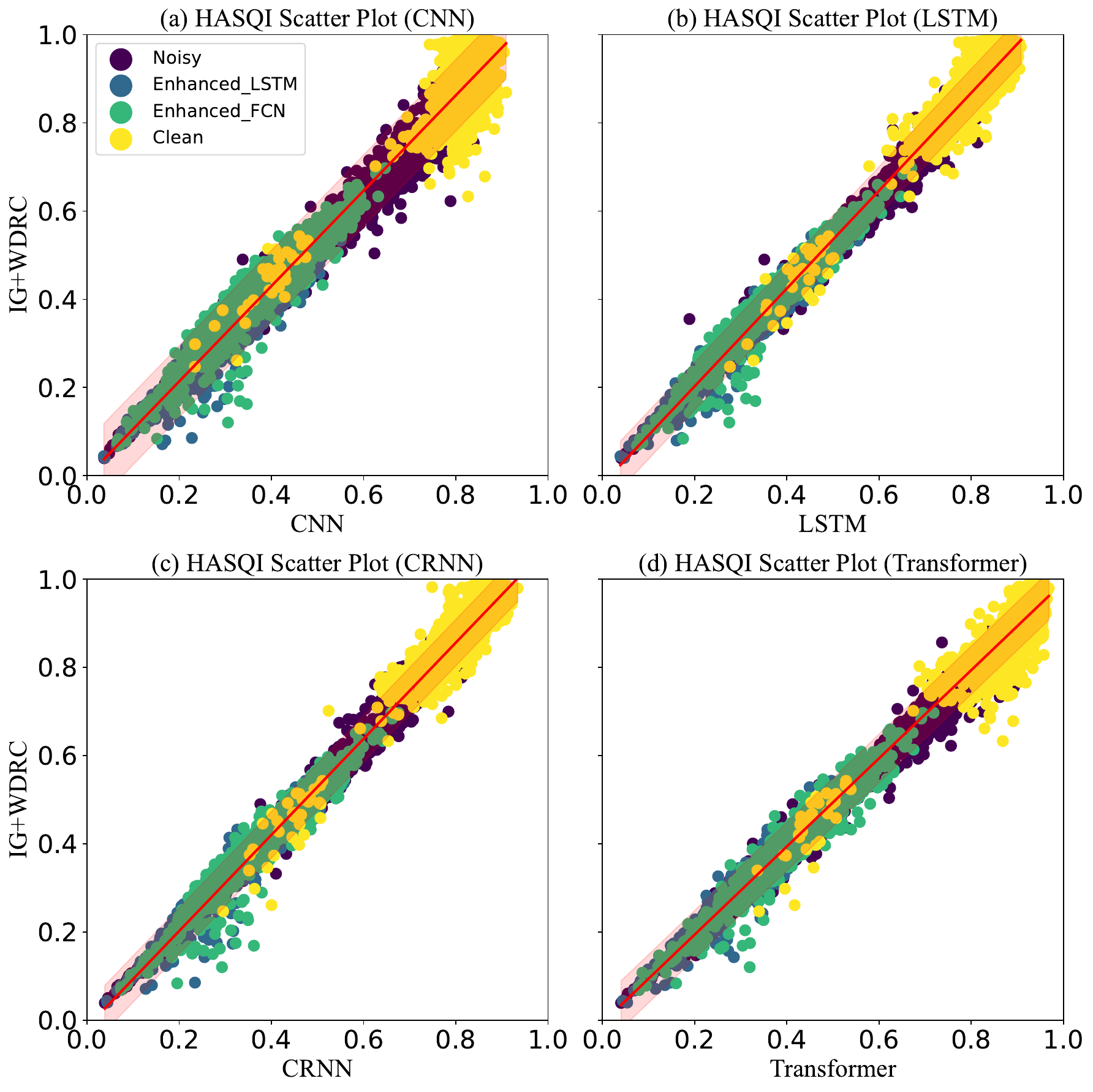}
\caption{VoiceBank-DEMAND dataset scatter plots for HASQI scores (a) CNN model architecture (b) LSTM model architecture (c) CRNN model architecture (d) Transformer model architecture}
\label{fig_9}
\end{figure}

\subsubsection{Scatter Plot Analysis}

We also utilize scatter plots to visually compare and assess the prediction distribution between the NeuroAMP and the ground truth (NAL-R+WDRC) for various speech datasets. Specifically, we selected two datasets: the TMHINT dataset, used during training, and the VoiceBank-DEMAND dataset, unseen during training. For evaluation, HASPI and HASQI scores are used as assessment metrics.

Fig. \ref{fig_6} displays scatter plots of the HASPI scores for the TMHINT dataset, separated by each model architecture: (a) CNN, (b) LSTM, (c) CRNN, and (d) Transformer. Fig. \ref{fig_7} shows scatter plots of the HASQI scores for the same dataset. Figures \ref{fig_8} and \ref{fig_9} illustrate scatter plots for HASPI and HASQI scores on the VoiceBank-DEMAND dataset, respectively, across all model architectures. In each scatter plot, a red regression line is plotted to indicate the best linear fit, and the shaded area around the line represents the 95\% confidence interval, providing a visual measure of the reliability of the linear relationship between the model predictions and the ground truth. Each plot is also color-coded to distinguish between different types of data: noisy, enhanced by FCN, enhanced by LSTM, and clean, allowing for a clear visualization of score distributions using HASPI and HASQI scores. The density of points along the diagonal in these plots reflects a high degree of agreement between the model outputs (NeuroAMP) and the ground truth (NAL-R+WDRC), suggesting superior model performance. Specifically, the Transformer and CRNN models exhibit distinctive patterns in these scatter plots. The Transformer model (Figures \ref{fig_6}d and \ref{fig_9}d) shows particularly dense clustering of points along the diagonal across all types of data, suggesting robust performance and effective handling of variations in audio quality. Moreover, the CRNN model (Figs 6c and 9c) shows better alignment with the ground truth but exhibits slightly more dispersion, which may be due to the CRNN’s varying ability to handle certain dataset characteristics or noise levels.

\subsubsection{Performance Evaluation on Music Dataset}

\textbf{Table V} presents a comprehensive evaluation of our models using the HAAQI metric to assess music quality against ground-truth scores across two test sets: the Cadenza Challenge and MUSDB18-HQ. LCC, SRCC, and MSE are reported, with statistically significant differences indicated. The Cadenza Challenge dataset, which is seen during the training phase, and the unseen MUSDB18-HQ dataset, which is not exposed to the model during training, are used to evaluate the models’ performance and abilities to generalize to new data.

\begin{table}[t]
  \centering
  \caption{Detailed evaluation results for Music datasets (mean $\pm$ std).}
  \label{tab:performance_music}
  \footnotesize
  \setlength{\tabcolsep}{2pt}
  \resizebox{\columnwidth}{!}{
    \begin{tabular}{lcccccc}
      \toprule
      \multirow{2}{*}{Model} & \multicolumn{3}{c}{Cadenza Challenge} & \multicolumn{3}{c}{MUSDB18-HQ (Unseen)} \\
      \cmidrule(lr){2-4}\cmidrule(lr){5-7}
      & LCC $\uparrow$ & SRCC $\uparrow$ & MSE $\downarrow$ & LCC $\uparrow$ & SRCC $\uparrow$ & MSE $\downarrow$ \\
      \midrule
      CNN         & 0.95 $\pm$ 0.01 & 0.93 $\pm$ 0.02 & 0.005 $\pm$ 0.001 & 0.97 $\pm$ 0.01 & 0.96 $\pm$ 0.01 & 0.003 $\pm$ 0.001 \\
      LSTM        & 0.97 $\pm$ 0.01 & 0.95 $\pm$ 0.01 & 0.007 $\pm$ 0.001 & 0.98 $\pm$ 0.00 & 0.98 $\pm$ 0.01 & 0.003 $\pm$ 0.001 \\
      CRNN        & 0.97 $\pm$ 0.01 & 0.96 $\pm$ 0.01 & 0.005 $\pm$ 0.001 & 0.98 $\pm$ 0.00 & 0.98 $\pm$ 0.01 & 0.002 $\pm$ 0.001 \\
    Transformer 
    & \textbf{0.98} $\pm$ 0.01 $^{* \dagger \ddagger}$
    & \textbf{0.97} $\pm$ 0.01 $^{* \dagger \ddagger}$
    & \textbf{0.003} $\pm$ 0.001 $^{* \dagger \ddagger}$
    & \textbf{0.99} $\pm$ 0.00 $^{* \dagger \ddagger}$
    & \textbf{0.99} $\pm$ 0.01 $^{* \dagger \ddagger}$
    & \textbf{0.002} $\pm$ 0.000 $^{* \dagger}$ \\
      \bottomrule
    \end{tabular}
  }
  \vspace{1mm}
  \begin{minipage}{\columnwidth}
    \raggedright
    Significance symbols for Transformer: $^*$ vs. CNN, $^\dagger$ vs. LSTM, $^\ddagger$ vs. CRNN $(p < 0.05)$.
  \end{minipage}
\end{table}

Notably, the performance metrics across these datasets demonstrate that our models, particularly the Transformer and CRNN architectures, can handle unseen scenarios effectively. This is evident from their notable performances in LCC, SRCC, and particularly low MSE values in the MUSDB18-HQ dataset. For example, the Transformer model achieves an LCC of 0.9888 and an SRCC of 0.9892 on the unseen MUSDB18-HQ dataset, indicating notable prediction performance and correlation with the ground truth, despite having no prior exposure during training. These findings confirm our models’ ability to perform robustly across different audio content types, such as speech and music, which inherently possess distinct acoustic characteristics. In addition, the models show good generalization performance on unseen data. Furthermore, the effectiveness of the models across both seen and unseen datasets highlights their potential for practical applications in various real-world scenarios.

\begin{figure}[!t]
\centering
\includegraphics[scale=0.29]{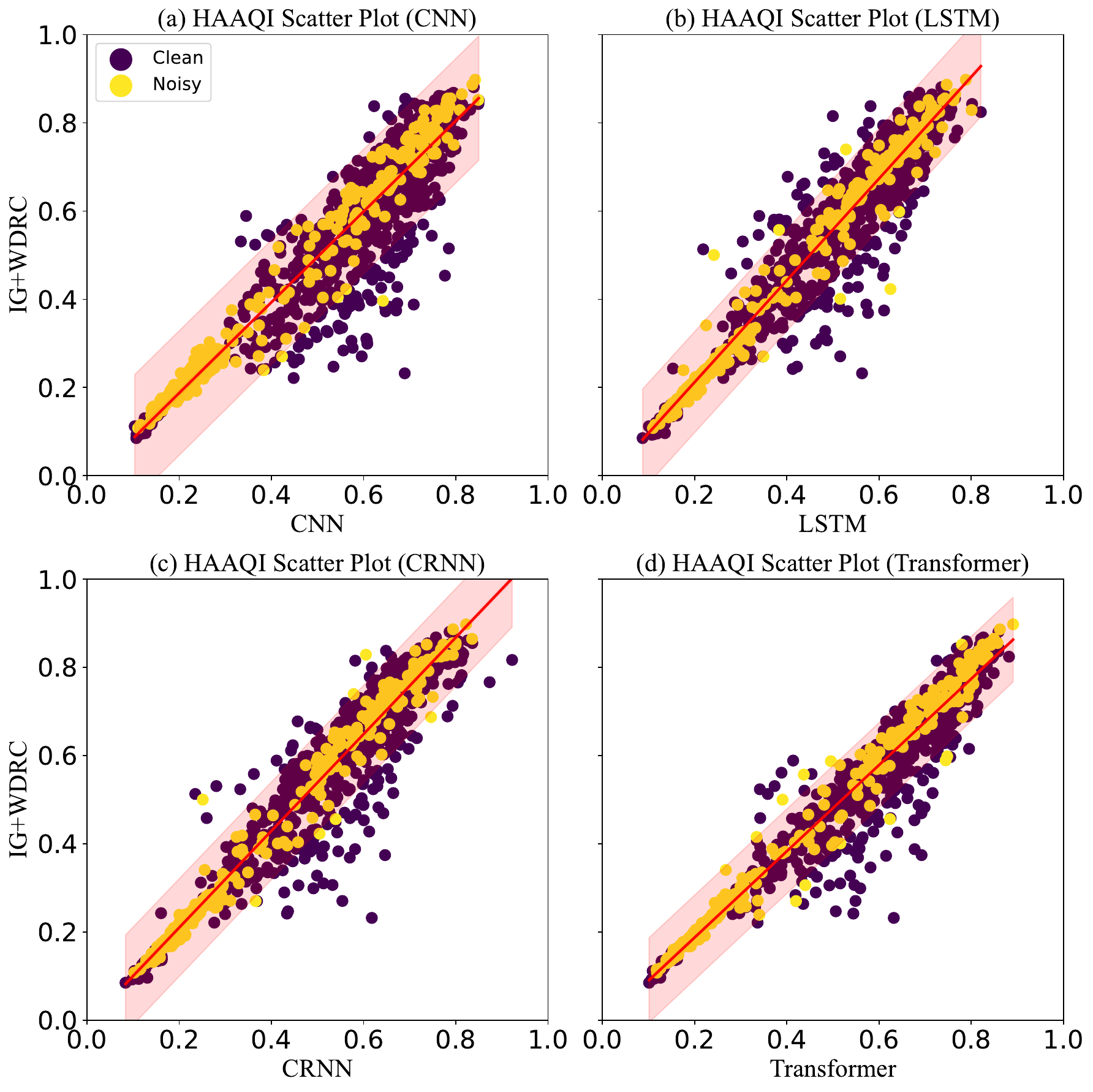}
\caption{ Clarity-Cadenza Challenge dataset scatter plots for HAAQI scores (a) CNN model architecture (b) LSTM model architecture (c) CRNN model architecture (d) Transformer model architecture}
\label{fig_10}
\end{figure}

\begin{figure}[!t]
\centering
\includegraphics[scale=0.29]{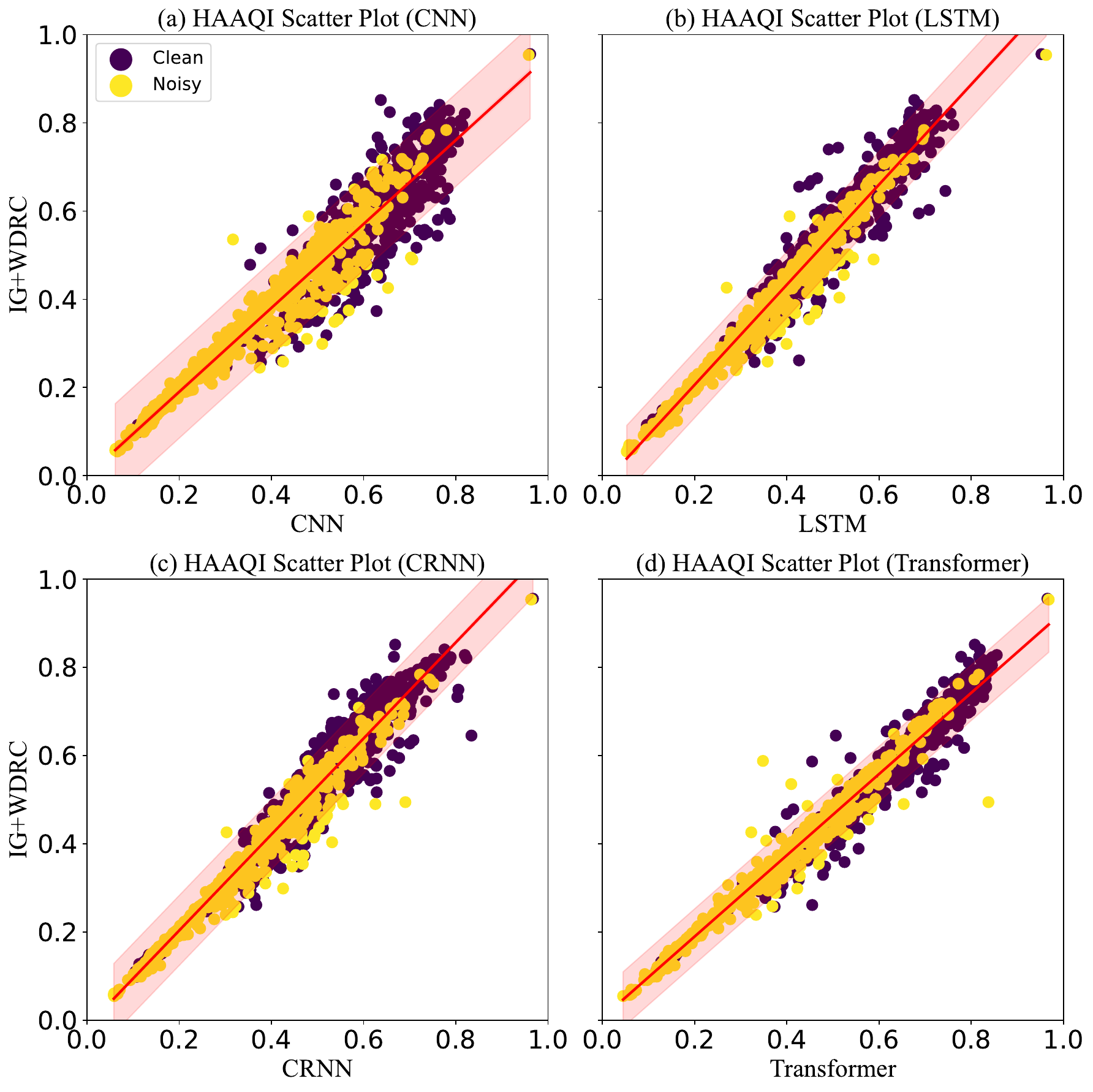}
\caption{MUSDB18-HQ dataset scatter plots for HAAQI scores (a) CNN model architecture (b) LSTM model architecture (c) CRNN model architecture (d) Transformer model architecture}
\label{fig_11}
\end{figure}

Next, we use scatter plots to visually analyze our models’ predictions of HAAQI scores across two different music datasets. In Fig. \ref{fig_10}, we show scatter plots for the Clarity-Cadenza Challenge dataset, showing the results for each model architecture: (a) CNN, (b) LSTM, (c) CRNN, and (d) Transformer. Fig. \ref{fig_11} presents scatter plots of HAAQI scores on the MUSDB18-HQ dataset, providing insights into the performance of the same models on this unseen music dataset. The plots indicate that all models performed well, with the Transformer architecture particularly excelling in handling both noisy and clean data, as evidenced by its densely clustered points along the diagonal line. These visualizations align with the numerical results in Table V and demonstrate the models' ability to maintain high accuracy and consistency across diverse scenarios, including seen and unseen datasets.

\subsubsection{Waveform Analysis}

\begin{figure}[!t]
\centering
\includegraphics[scale=0.28]{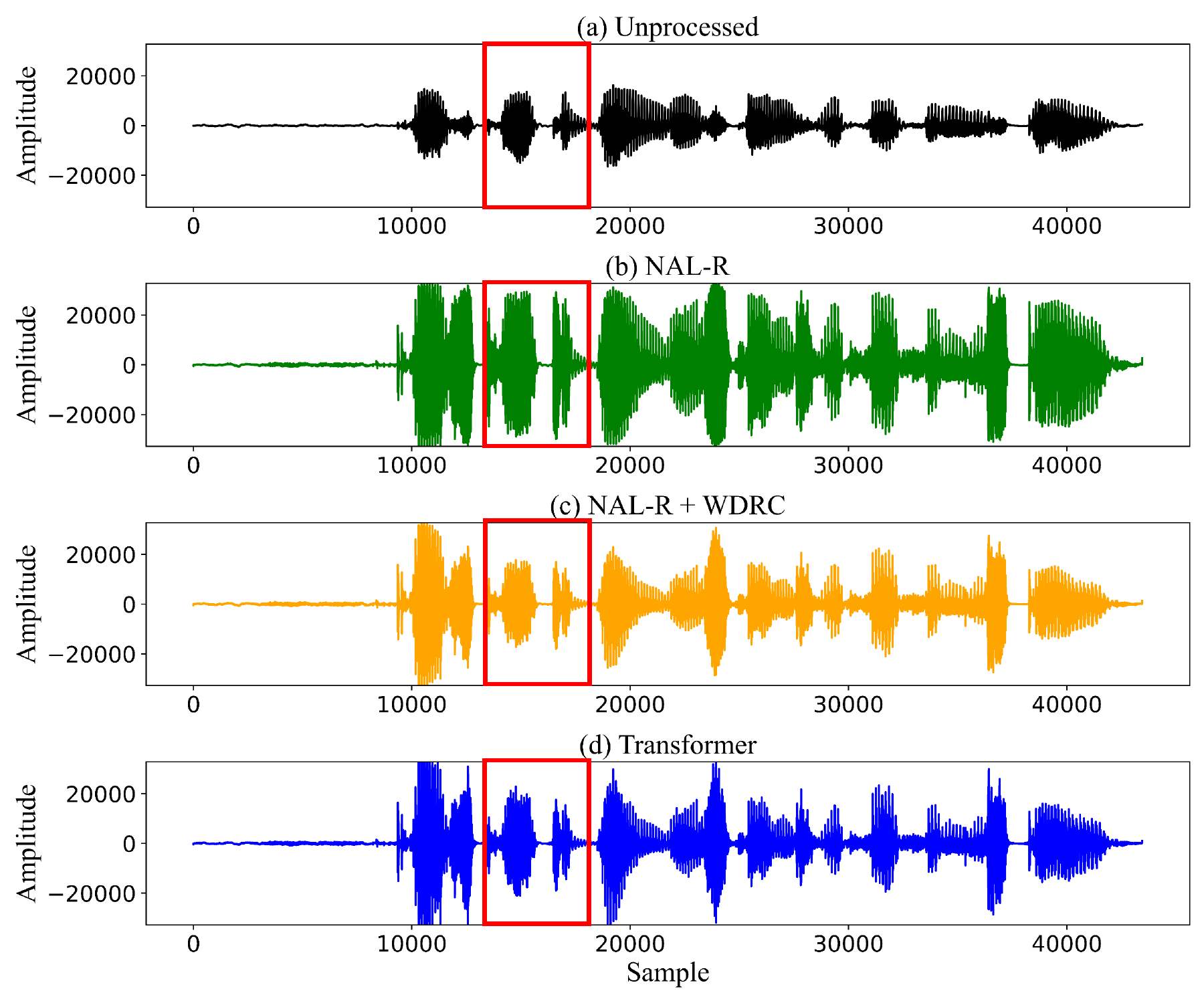}
\caption{Waveform comparison: (a) unprocessed, (b) NAL-R, (c) NAL-R+WDRC, (d) NeuroAMP(Transformer)}
\label{waveform_fig}
\end{figure}

To further validate our model's performance, we conducted an in-depth analysis encompassing waveform comparison, spectral analysis, and frequency band energy over time. Using a moderate-severity sloping audiogram with high-frequency hearing loss as the listener profile, we compared the unprocessed NAL-R, NAL-R+WDRC, and NeuroAMP amplified signals. Fig. \ref{waveform_fig} presents the waveform comparison, revealing that the Transformer model's output closely aligns with the ground-truth (NAL-R+WDRC) signal. The similarity in waveform patterns demonstrates our model’s effectiveness in accurately replicating the ground-truth waveform, as indicated by the highlighted regions (red boxes).

\subsubsection{Spectrogram Analysis}

\begin{figure}[!t]
\centering
\includegraphics[scale=0.30]{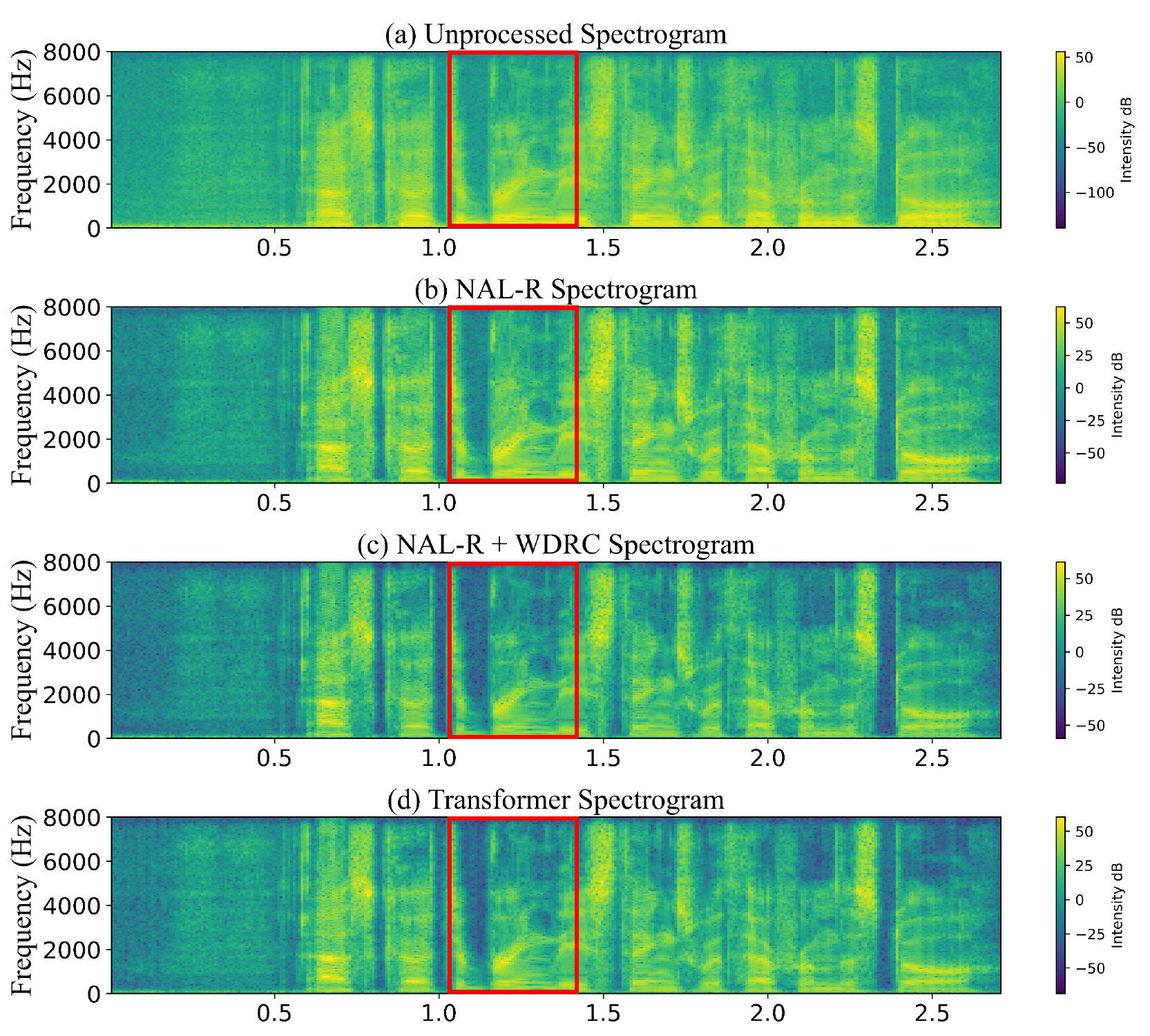}
\caption{Spectrogram comparison: (a) unprocessed, (b) NAL-R, (c) NAL-R+WDRC, (d) NeuroAMP (Transformer)}
\label{specc}
\end{figure}

Fig. \ref{specc} presents the spectral comparison for unprocessed NAL-R, NAL-R+WDRC, and NeuroAMP amplified signals. The spectral analysis aligns with the findings from the waveform comparison, indicating that the Transformer model's spectral characteristics closely resemble those of the NAL-R+WDRC amplified signal. This consistency across both waveform and spectral domains highlights NeuroAMP's capability to replicate the performance characteristics of the ground-truth signals.

\subsubsection{Frequency Band Analysis}

\begin{figure}[!t]
\centering
\includegraphics[scale=0.23]{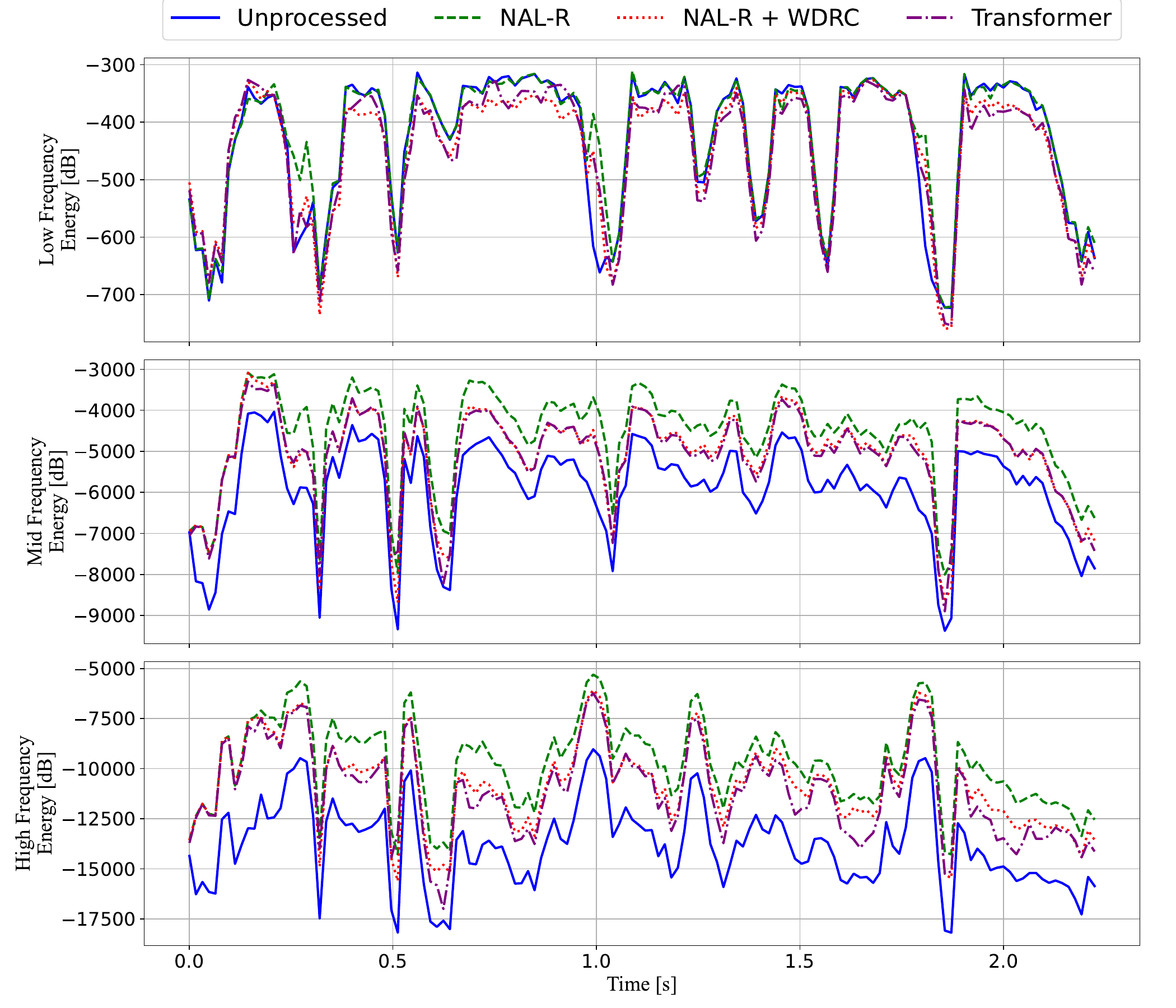}
\caption{Frequency band energy over time comparison for unprocessed, NAL-R, NAL-R+WDRC, and NeuroAMP (Transformer). (Top panel): low frequency; (Middle panel): mid. frequency; (Bottom panel): high frequency.}
\label{band_energy}
\end{figure}

Fig. \ref{band_energy} illustrates the comparative analysis of frequency band energy over time for unprocessed, NAL-R, NAL-R+WDRC, and NeuroAMP-processed signals. This analysis provides insights into how each processing method affects the time-frequency distribution, focusing on specific frequency bands and their energy distribution. In the low-frequency band (0–500 Hz), all signals show similar overall energy trends. NAL-R and NAL-R+WDRC maintain comparable low-frequency energy, with the latter adding subtle compression-induced smoothing. Notably, the NeuroAMP-processed signal closely mirrors the energy pattern of the NAL-R+WDRC, demonstrating its ability to effectively replicate low-frequency signal processing.

In the mid-frequency band (500–2000 Hz), the unprocessed signal shows the lowest energy levels. The NAL-R processing significantly increases energy in this range, enhancing mid-frequency components crucial for speech intelligibility. While similar to NAL-R in energy levels, the NAL-R+WDRC processing introduces slight variations due to dynamic range compression. The NeuroAMP model closely aligns with the NAL-R+WDRC energy distribution, demonstrating high accuracy in replicating mid-frequency amplification and compression.

In the high-frequency band (2000-8000 Hz), the unprocessed signal maintains low energy levels. The NAL-R processed signal shows significant amplification, amplifying high-frequency components crucial for speech intelligibility. As with other bands, NAL-R+WDRC processing introduces minor variations due to compression effects. The NeuroAMP-processed signal closely mirrors the NAL-R+WDRC pattern, effectively replicating both high-frequency amplification and compression.

Overall, this detailed analysis across specific frequency bands confirms that the NeuroAMP model performs comparably to traditional NAL-R and NAL-R+WDRC processing techniques. By effectively preserving the time-frequency energy distribution, the NeuroAMP model closely mirrors the ground truth provided by the NAL-R+WDRC processed signal.

\begin{figure}[!t]
\centering
\includegraphics[scale=0.24]{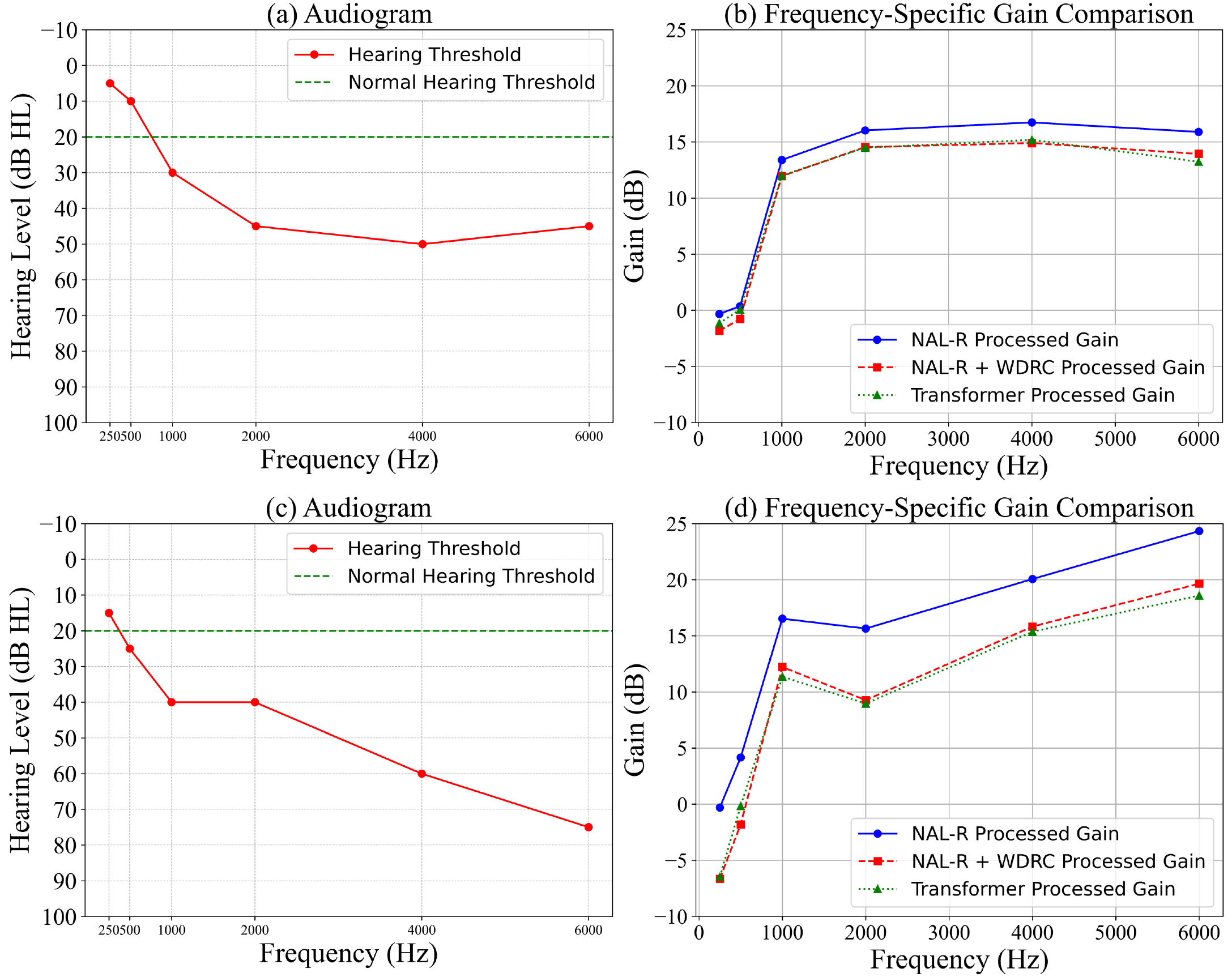}
\caption{Frequency-specific gain comparison of NerouAMP with NAL-R and NAL-R+WDRC.}
\label{gain}
\end{figure}

\subsubsection{Gain Function Analysis}

Additionally, in our comparative study, we analyze gain patterns across different listener audiograms to evaluate the performance of various signal processing methods. Fig. \ref{gain} presents an analysis of audiograms and frequency-specific gain comparisons for two subjects with different hearing loss profiles. Panels (a) and (c) show the audiograms, illustrating the hearing thresholds of the subjects relative to the normal hearing threshold. Panels (b) and (d) display the frequency-specific gain comparisons for the NAL-R, NAL-R+WDRC, and NeuroAMP-processed signals.

In panel (a), the audiogram displays the hearing threshold (red line) for the first subject, indicating significant hearing loss at higher frequencies. The normal hearing threshold (green dashed line) serves as a reference, highlighting the extent of hearing loss. Panel (c) shows the audiogram for the second subject, who has more pronounced hearing loss, particularly at higher frequencies, compared to the first subject. Panels (b) and (d) compare the gains provided by different processing methods. In Panel (b), the NAL-R processed gain (blue line) shows substantial amplification at lower frequencies with a gradual decrease towards higher frequencies. The NAL-R+WDRC processed gain (red line) closely follows the NAL-R curve but exhibits a more stable gain distribution across frequencies due to compression effects. The NeuroAMP-processed gain (green line with triangles) closely mirrors the NAL-R+WDRC gain, demonstrating NeuroAMP's effective replication of traditional processing methods.

In panel (d), the NAL-R processed gain (blue line) shows a more pronounced increase, particularly at higher frequencies, to address the greater hearing loss shown in Panel (c). The NAL-R+WDRC processed gain (red line) continues to follow the NAL-R curve, with compression effects providing a stable gain distribution. The NeuroAMP-processed gain (green line with triangles) closely aligns with the NAL-R+WDRC gain pattern, demonstrating NeuroAMP’s effective replication of the gain characteristics. This analysis highlights that the NeuroAMP model successfully replicates the gain features of NAL-R+WDRC processing across both hearing loss profiles. Its ability to provide appropriate amplification and compression tailored to specific hearing loss profiles indicates that it is a viable alternative for hearing aid signal processing. By matching the performance of established methods, the NeuroAMP model ensures comparable speech intelligibility and listening comfort for users, similar to traditional methods.

\subsection{Denoising NeuroAMP Performance Evaluation}

This section evaluates the performance of the Denoising NeuroAMP model, which combines speech enhancement and personalized amplification in a single unified model. The evaluation was conducted on the VoiceBank-DEMAND dataset. We compare Denoising NeuroAMP to two baselines:

\begin{itemize}
\item NAL-R+WDRC: This represents the conventional approach of applying the NAL-R prescription formula and WDRC directly to the noisy speech signal.
\item Denoising$\to$NAL-R+WDRC: This is a two-stage baseline that first uses a separate LSTM-based speech enhancement model to enhance noisy speech, and then uses NAL-R+WDRC for amplification.
\end{itemize}

Table VI lists the HASPI and HASQI scores of Denoising NeuroAMP and the two baselines on the noisy test set of the VoiceBank-DEMAND dataset.

\begin{table}[t]
    \centering
    \caption{Performance Comparison of Denoising NeuroAMP on the VoiceBank-DEMAND Noisy Test Set}
    \label{tab:haspi_hasqi_comparison}
    \resizebox{\columnwidth}{!}{
    \begin{tabular}{lcc}
        \toprule
        Method & HASPI & HASQI \\
        \midrule
        NAL-R+WDRC (Noisy Input) & 0.89 & 0.55 \\
        Denoising$\to$NAL-R+WDRC (Two-Stage) & 0.85 & 0.30 \\
        Denoising NeuroAMP (End-to-End) & 0.90 & 0.59 \\
        \bottomrule
    \end{tabular}
    }
\end{table}

The results show that Denoising NeuroAMP outperforms both baselines in both HASPI and HASQI scores. Specifically, Denoising NeuroAMP achieves a HASPI score of 0.90 and a HASQI score of 0.59, while NAL-R+WDRC has scores of 0.89 and 0.55, and the two-stage Denoising$\to$NAL-R+WDRC baseline has scores of 0.85 and 0.30. The improvements achieved by Denoising NeuroAMP compared to the two-stage Denoising$\to$NAL-R+WDRC baseline are statistically significant (paired t-test, p $<$ 0.05). These findings suggest that Denoising NeuroAMP, an integrated approach that jointly optimizes speech enhancement and personalized amplification, is more effective than applying these processes independently. The lower performance of the two-stage baseline, particularly with respect to HASQI, highlights potential drawbacks of the two-stage approach, as errors from the initial enhancement stage may negatively affect subsequent amplification.
The superior performance of Denoising NeuroAMP confirms its potential to improve speech intelligibility and quality for hearing aid users in noisy environments. By learning a unified denoising and amplification process tailored to an individual's hearing profile, Denoising NeuroAMP provides a promising alternative approach for enhancing hearing aid technology.

\subsection{Subjective Evaluations}

\subsubsection{Perceptual Similarity Analysis}

This section presents a preference test comparing our proposed NeuroAMP model with the traditional NAL-R+WDRC processing. The evaluation involved 20 subjects, evenly split between males and females, all confirmed to have normal hearing through audiometric testing \footnote{All ethical and experimental procedures were approved by the Institutional Review Board (IRB) on Biomedical Science Research at Academia Sinica under protocol number AS-IRB-BM-24026. All experiments were conducted in accordance with the approved IRB protocols}. To simulate hearing loss, we used the MSBG hearing loss simulator~\cite{msbg}, enabling a realistic assessment of processed audio as perceived by individuals with hearing loss. Each subject listened to pairs of audio samples processed by either the NAL-R+WDRC or NeuroAMP models, which were then further processed using the MSBG hearing loss simulator. They were asked to determine whether sample A was better, sample B was better, or if the samples were similar and indistinguishable. The sample order was randomized, and the tests were conducted in a double-blind manner. We selected a diverse set of audio samples from our test dataset to ensure a comprehensive evaluation, including 10 clean speech samples, 10 noisy speech samples, 10 speech samples enhanced by the FCN model, 10 speech samples enhanced by the LSTM model, 5 clean music samples, and 5 noisy music samples. Participants evaluated each pair based on intelligibility (clarity of speech) and quality (overall audio quality, including naturalness and pleasantness).

\begin{figure}[!t]
\centering
\includegraphics[scale=0.35]{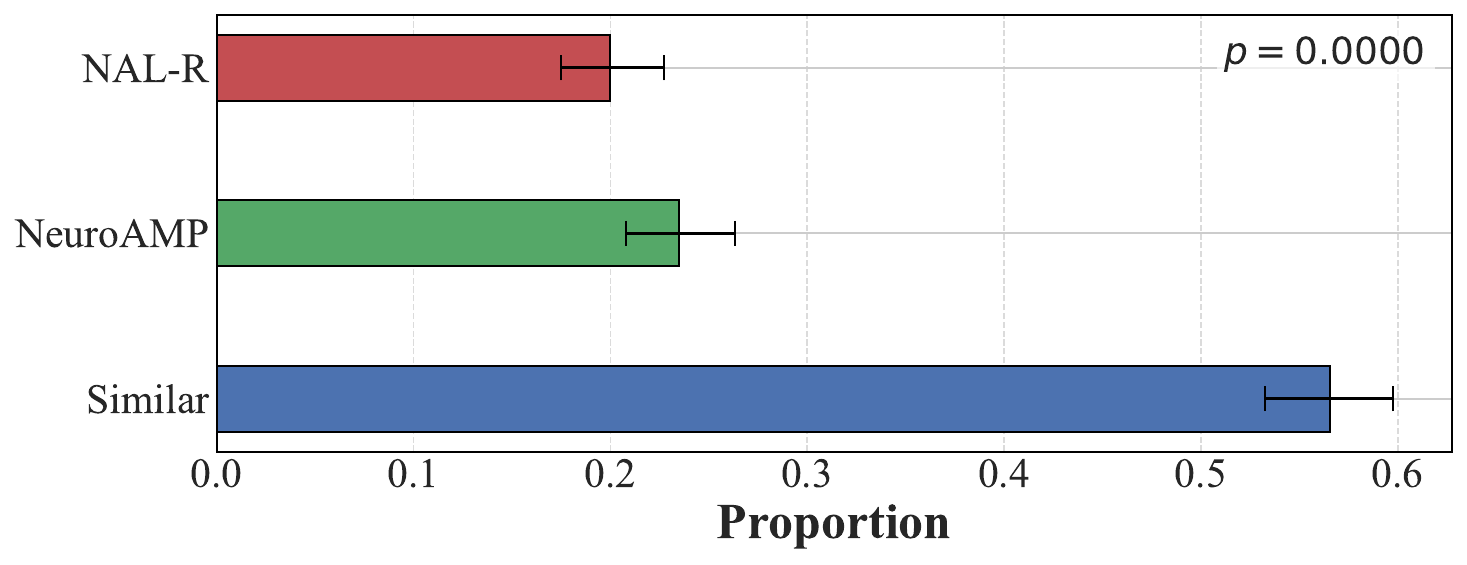}
\caption{Aggregate results showing participant preferences for NeuroAMP, NAL-R+WDRC, or if both types were considered similar, across all sample types.}
\label{subjective1}
\end{figure}

\begin{figure}[!t]
\centering
\includegraphics[scale=0.35]{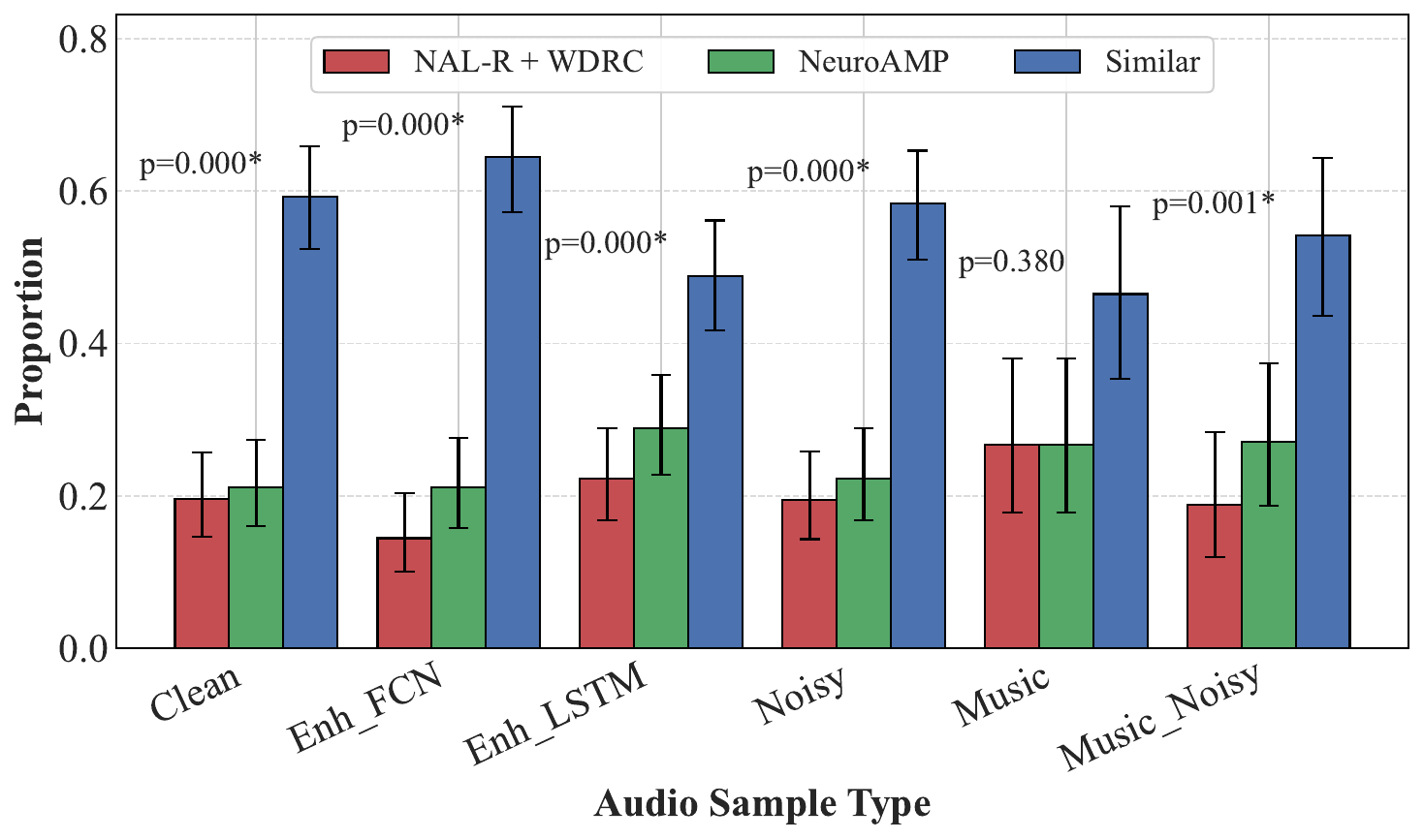}
\caption{Detailed breakdown showing preferences for NeuroAMP, NAL-R+WDRC, or if both types were considered similar for each audio sample category.}
\label{subjective2}
\end{figure}

Fig. \ref{subjective1} shows the evaluation results, aggregating preferences and ratings across different sample types. Proportions for each response category are shown with 95\% confidence intervals, and statistical significance was assessed using a chi-square goodness-of-fit test, with the resulting p-value annotated in the figure. Detailed visual representations of the evaluation results for each sample type—clean speech, noisy speech, enhanced speech by FCN, enhanced speech by LSTM, clean music, and noisy music—are shown in Fig. \ref{subjective2}. As shown in Fig. \ref{subjective1}, responses indicating 'Similar' significantly outnumbered preferences for either 'NAL-R+WDRC' or 'NeuroAMP'. Fig. \ref{subjective2} shows a consistent trend: "Similar" decisions prevail across all conditions, including speech, noisy speech, enhanced speech, and music scenarios. These results confirm that the proposed NeuroAMP model effectively replicates the perceptual characteristics of NAL-R+WDRC, demonstrating the viability of a deep learning-based amplification approach.

\subsubsection{Mean Opinion Score (MOS)}

\begin{figure}[!t]
\centering
\includegraphics[scale=0.34]{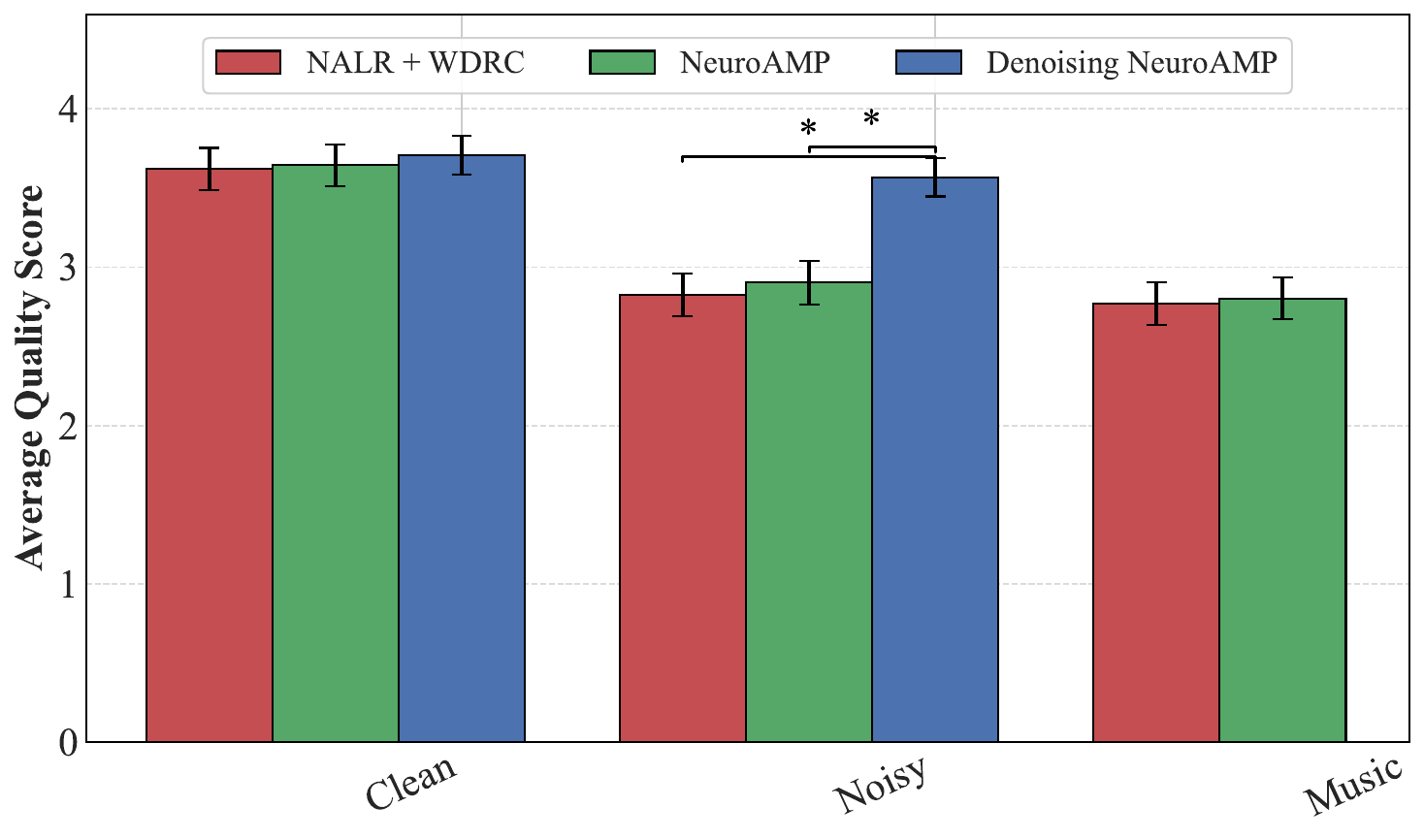}
\caption{MOS comparison for perceived audio quality across three audio categories (clean speech, noisy speech, and music) processed using NAL-R+WDRC, NeuroAMP, and Denoising NeuroAMP methods. Bars represent mean scores with 95\% confidence intervals. Significant differences between processing methods (p~$<$~0.05) are marked by asterisks (*). The hearing-impaired speech signals were simulated using the MSBG model.}
\label{mos}
\end{figure}

To quantitatively assess the perceived audio quality of the processed samples, we conducted a MOS evaluation involving 30 listeners. The audio samples were processed using three amplification approaches: the traditional NAL-R+WDRC processing, the proposed NeuroAMP model, and Denoising NeuroAMP. Subsequently, each audio sample was processed using the MSBG hearing loss simulator to emulate auditory experiences representative of individuals with hearing impairment. The listeners were asked to evaluate audio samples from three distinct categories: clean speech, noisy speech, and music. Each participant rated perceived audio quality using a MOS scale from 1 ("poor" quality) to 5 ("excellent" quality). The evaluation results, including average scores and 95\% confidence intervals, are presented in Fig.~\ref{mos}. Statistical significance was assessed using one-way ANOVA and pairwise t-tests, with significant differences indicated by asterisks (*) above the bars.

As shown in Fig.~\ref{mos}, the three processing approaches demonstrate comparable performance for clean speech, with overlapping confidence intervals. Specifically, Denoising NeuroAMP achieved the highest MOS (3.71), closely followed by NeuroAMP (3.64) and the conventional NAL-R+WDRC method (3.62). In noisy speech conditions, Denoising NeuroAMP achieves the best performance with an average MOS of 3.57, and the pairwise comparison shows this difference to be statistically significant compared to the other methods (as indicated by asterisks). NeuroAMP and NAL-R+WDRC also show comparable performance, with NeuroAMP achieving a higher MOS than NAL-R+WDRC (2.90 vs. 2.82). For music, NeuroAMP and NAL-R+WDRC produced closely matched results, with NeuroAMP slightly outperforming NAL-R+WDRC (2.80 vs. 2.77).

Overall, the MOS evaluation confirms that the proposed NeuroAMP model and the conventional NAL-R+WDRC processing offer similar perceptual quality across various audio categories. Notably, the addition of denoising capabilities through Denoising NeuroAMP significantly improves audio quality in noisy speech scenarios, highlighting the potential to further enhance deep learning-based amplification frameworks.

\begin{figure}[!t]
\centering
\includegraphics[scale=0.34]{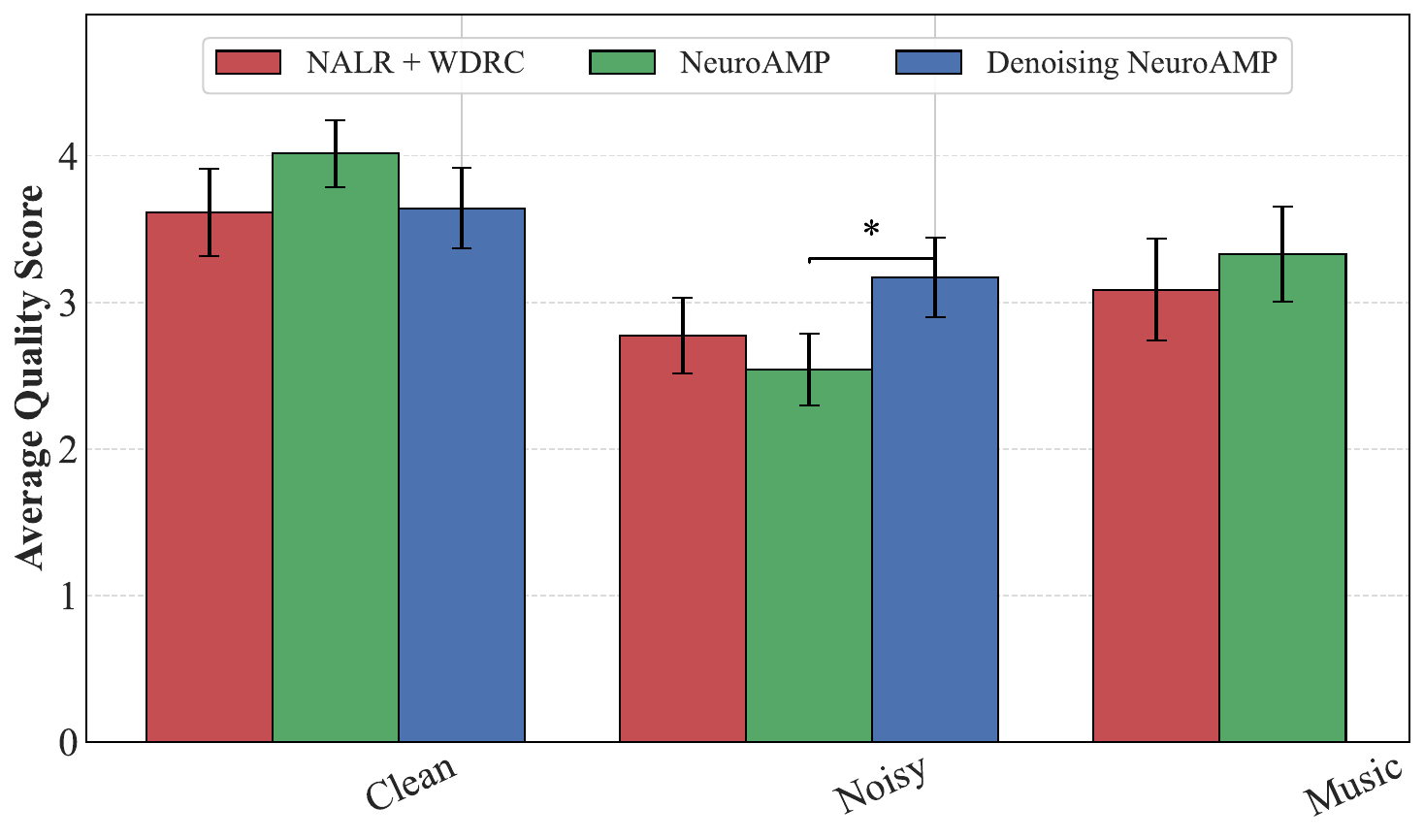}
\caption{MOS comparison for perceived audio quality across three audio categories (clean speech, noisy speech, and music) processed using NAL-R+WDRC, NeuroAMP, and Denoising NeuroAMP methods. Seven subjects with hearing loss participated in the listening test.}
\label{mos_HA}
\end{figure}

Finally, we recruited 7 individuals with hearing loss to participate in the MOS listening test. Each subject had a specific audiogram, and the evaluation employed the same set of speech samples as used in Fig.\ref{mos}. Prior to the listening tests, audiograms for each participant's target ear were obtained. Based on these audiograms, the speech signals were processed using NAL-R+WDRC, NeuroAMP, and Denoising NeuroAMP, and then presented to the participants without the use of hearing aids. The resulting MOS scores (mean ± 95\% confidence intervals) are shown in Fig.~\ref{mos_HA}. Statistical significance was assessed using one-way ANOVA and Bonferroni-corrected pairwise t-tests, with significant differences marked by asterisks (*) in the figure. The trends observed in Fig. \ref{mos_HA} align with those in Fig. \ref{mos}. In clean and music conditions, NAL-R+WDRC and NeuroAMP demonstrated comparable performance, with NeuroAMP slightly outperforming NAL-R+WDRC. Under noisy conditions, Denoising NeuroAMP exhibited significant improvements over the other two methods.

\section{Conclusion}
\label{sec:conclusion}

This research introduces NeuroAMP, a novel deep learning-based system for personalized hearing aid amplification, and Denoising NeuroAMP, an extension that integrates noise reduction for better performance in adverse environments. We investigated four neural network architectures: CNN, LSTM, CRNN, and Transformer, using spectral features and audiograms as inputs. Extensive evaluation using HASPI, HASQI, and HAAQI, together with statistical measures like LCC, SRCC, and MSE, demonstrated the superior performance of the Transformer within the NeuroAMP framework. Notably, NeuroAMP achieves SRCC scores of 0.9810 (HASQI) and 0.9914 (HASPI) on the unseen VoiceBank-DEMAND dataset, and 0.9892 (HAAQI) on the MUSDB18-HQ dataset, validating the effectiveness of our data augmentation strategy. Subjective listening tests further confirmed that NeuroAMP produces outputs perceptually similar to traditional methods while providing greater adaptability. Furthermore, Denoising NeuroAMP outperformed conventional NAL-R+WDRC and two-stage baselines on the VoiceBank-DEMAND dataset, highlighting its potential for enhanced speech intelligibility in noisy environments. In conclusion, this work demonstrates the potential of NeuroAMP and Denoising NeuroAMP as personalized, user-centric solutions for amplification and noise reduction. In our future work, we will focus on refining these models, exploring real-time implementation, incorporating user feedback, and integrating other essential modules, such as a hearing loss model for compensation, to achieve improved personalized hearing assistance solutions.

\bibliographystyle{IEEEbib}
\bibliography{ref}

\begin{thebibliography}{10}

\bibitem{shukla2020hearing}
A.~Shukla, M.~Harper, E.~Pedersen, A.~Goman, J.~J. Suen, C.~Price, J.~Applebaum, M.~Hoyer, F.~R. Lin, and N.~S. Reed,
\newblock ``Hearing loss, loneliness, and social isolation: A systematic review,''
\newblock {\em Otolaryngology-Head and Neck Surgery: Official Journal of American Academy of Otolaryngology-Head and Neck Surgery}, vol. 162, no. 5, pp. 622--633, 2020.

\bibitem{lawrence2020hearing}
B.~J. Lawrence, D.~M.~P. Jayakody, R.~J. Bennett, R.~H. Eikelboom, N.~Gasson, and P.~L. Friedland,
\newblock ``Hearing loss and depression in older adults: a systematic review and meta-analysis,''
\newblock {\em The Gerontologist}, vol. 60, no. 3, pp. 137--154, 2020.

\bibitem{ciorba2012impact}
A.~Ciorba, C.~Bianchini, S.~Pelucchi, and A.~Pastore,
\newblock ``The impact of hearing loss on the quality of life of elderly adults,''
\newblock {\em Clinical Interventions in Aging}, pp. 159--163, 2012.

\bibitem{lin2013hearing}
F.~R. Lin, K.~Yaffe, J.~Xia, Q.-L. Xue, T.~B. Harris, E.~Purchase-Helzner, S.~Satterfield, H.~N. Ayonayon, L.~Ferrucci, and E.~M. Simonsick,
\newblock ``Hearing loss and cognitive decline in older adults,''
\newblock {\em JAMA Internal Medicine}, vol. 173, no. 4, pp. 293--299, 2013.

\bibitem{WHO_hearing_loss_2024}
{World Health Organization},
\newblock ``Deafness and hearing loss,'' \url{https://www.who.int/news-room/fact-sheets/detail/deafness-and-hearing-loss}, Feb. 2024.

\bibitem{Audiologists_org_2024}
{Audiologists.org},
\newblock ``Deafness and hearing loss statistics 2024,'' \url{https://www.audiologists.org/deafness-and-hearing-loss-statistics-2024}, 2024.

\bibitem{HA2018use}
{Centers for Disease Control and Prevention},
\newblock ``National center for health statistics. national health and nutrition examination survey: {NHANES} 2015--2016 questionnaire data,'' 2018.

\bibitem{HA2018use2}
W.~Chien and F.~R. Lin,
\newblock ``Prevalence of hearing aid use among older adults in the united states,''
\newblock {\em Archives of Internal Medicine}, vol. 172, no. 3, pp. 292--293, 2012.

\bibitem{HA2015use3}
A.~McCormack and H.~Fortnum,
\newblock ``Why do people fitted with hearing aids not wear them?,''
\newblock {\em International Journal of Audiology}, vol. 52, no. 5, pp. 360--368, 2013.

\bibitem{NAL-R}
D.~Byrne and H.~Dillon,
\newblock ``The national acoustic laboratories' ({NAL}) new procedure for selecting the gain and frequency response of a hearing aid,''
\newblock {\em Ear and Hearing}, vol. 7, no. 4, pp. 257--265, 1986.

\bibitem{DSL}
S.~Scollie, R.~Seewald, L.~Cornelisse, S.~Moodie, M.~Bagatto, D.~Laurnagaray, S.~Beaulac, and J.~Pumford,
\newblock ``The desired sensation level multistage input/output algorithm,''
\newblock {\em Trends in Amplification}, vol. 9, no. 4, pp. 159--197, 2005.

\bibitem{NAL-NL1}
D.~Byrne, H.~Dillon, T.~Ching, R.~Katsch, and G.~Keidser,
\newblock ``Nal-{NL}1 procedure for fitting nonlinear hearing aids: characteristics and comparisons with other procedures,''
\newblock {\em Journal of the American Academy of Audiology}, vol. 12, no. 1, pp. 37--51, 2001.

\bibitem{Compression}
P.~E. Souza,
\newblock ``Effects of compression on speech acoustics, intelligibility, and sound quality,''
\newblock {\em Trends in Amplification}, vol. 6, no. 4, pp. 131--165, 2002.

\bibitem{WDRCbenefit1}
L.~M. Jenstad, R.~C. Seewald, L.~E. Cornelisse, and J.~Shantz,
\newblock ``Comparison of linear gain and wide dynamic range compression hearing aid circuits: aided speech perception measures,''
\newblock {\em Ear and Hearing}, vol. 20, no. 2, pp. 117--126, 1999.

\bibitem{NAL-NL2}
G.~Keidser, H.~Dillon, M.~Flax, T.~Ching, and S.~Brewer,
\newblock ``The {NAL}-{NL}2 prescription procedure,''
\newblock {\em Audiology Research}, vol. 1, no. 1, pp. e24, 2011.

\bibitem{DSLm}
R.~Seewald, S.~Moodie, S.~Scollie, and M.~Bagatto,
\newblock ``The {DSL} method for pediatric hearing instrument fitting: historical perspective and current issues,''
\newblock {\em Trends in Amplification}, vol. 9, no. 4, pp. 145--157, 2005.

\bibitem{OpenMHA}
Hendrik Kayser, Tobias Herzke, Paul Maanen, Max Zimmermann, Giso Grimm, and Volker Hohmann,
\newblock ``Open community platform for hearing aid algorithm research: open master hearing aid (openmha),''
\newblock {\em SoftwareX}, vol. 17, pp. 100953, 2022.

\bibitem{wang2017deep}
D.~Wang,
\newblock ``Deep learning reinvents the hearing aid,''
\newblock {\em IEEE spectrum}, vol. 54, no. 3, pp. 32--37, 2017.

\bibitem{akeroyd20232nd}
M.~A. Akeroyd, W.~Bailey, J.~Barker, T.~J. Cox, J.~F. Culling, S.~Graetzer, G.~Naylor, Z.~Podwinska, and Z.~Tu,
\newblock ``The 2nd clarity enhancement challenge for hearing aid speech intelligibility enhancement: Overview and outcomes,''
\newblock in {\em Proc. IEEE International Conference on Acoustics, Speech and Signal Processing (ICASSP)}, 2023, pp. 1--5.

\bibitem{tu22_interspeech}
Z.~Tu, N.~Ma, and J.~Barker,
\newblock ``Exploiting hidden representations from a {DNN}-based speech recogniser for speech intelligibility prediction in hearing-impaired listeners,''
\newblock in {\em Proc. Interspeech}, 2022, pp. 3488--3492.

\bibitem{edozezario22_interspeech}
R.~E. Zezario, F.~Chen, C.-S. Fuh, H.-M. Wang, and Y.~Tsao,
\newblock ``{MBI-Net}: A non-intrusive multi-branched speech intelligibility prediction model for hearing aids,''
\newblock in {\em Proc. Interspeech}, 2022, pp. 3944--3948.

\bibitem{MAWALIM2023109663}
C.~O. Mawalim, B.~A. Titalim, S.~Okada, and M.~Unoki,
\newblock ``Non-intrusive speech intelligibility prediction using an auditory periphery model with hearing loss,''
\newblock {\em Applied Acoustics}, vol. 214, pp. 109663, 2023.

\bibitem{timit}
J.~S. Garofolo, L.~Lamel, W.~M. Fisher, J.~G. Fiscus, and D.~S. Pallett,
\newblock ``{DARPA} {TIMIT} acoustic-phonetic continuous speech corpus {CD-ROM}. {NIST} speech disc 1-1.1,''
\newblock {\em NASA STI/Recon Technical Report N}, vol. 93, pp. 27403, 1993.

\bibitem{cedenza1}
G.~R. Dabike, S.~Bannister, J.~Firth, S.~Graetzer, R.~Vos, M.~A. Akeroyd, and W.~Whitmer,
\newblock ``The first cadenza signal processing challenge: Improving music for those with a hearing loss,'' 2023.

\bibitem{vctk}
C.~Valentini-Botinhao,
\newblock ``Noisy speech database for training speech enhancement algorithms and {TTS} models,'' 2016,
\newblock [Dataset]. University of Edinburgh, School of Informatics, Centre for Speech Technology Research (CSTR).

\bibitem{musdb}
Z.~Rafii, A.~Liutkus, F.~St{\"o}ter, S.~I. Mimilakis, and R.~M. Bittner,
\newblock ``{MUSDB18-HQ} -- an uncompressed version of {MUSDB18},'' 2019.

\bibitem{NAL-RP}
D.~Byrne, A.~Parkinson, and P.~Newall,
\newblock ``Modified hearing aid selection procedures for severe/profound hearing losses,''
\newblock in {\em The Vanderbilt Hearing Aid Report II}, G.~Studebaker, F.~Bess, and L.~Beck, Eds. York Press, 1991.

\bibitem{WDRCblock}
T.~May, B.~Kowalewski, and T.~Dau,
\newblock {\em Scene-aware dynamic-range compression in hearing aids}, pp. 763--799,
\newblock Springer International Publishing, 2020.

\bibitem{speaker_emb}
X.~Ji, M.~Yu, C.~Zhang, D.~Su, T.~Yu, X.~Liu, and D.~Yu,
\newblock ``Speaker-aware target speaker enhancement by jointly learning with speaker embedding extraction,''
\newblock in {\em Proc. {IEEE} International Conference on Acoustics, Speech and Signal Processing. ({ICASSP})}, 2020, pp. 7294--7298.

\bibitem{fcnse}
S.-W. Fu, Y.~Tsao, X.~Lu, and H.~Kawai,
\newblock ``Raw waveform-based speech enhancement by fully convolutional networks,''
\newblock in {\em Proc. {APSIPA} Annual Summit and Conference}, Kuala Lumpur, Malaysia, 2017.

\bibitem{cnn_speech}
H.~S. Mashiana, A.~Salaria, and K.~Kaur,
\newblock ``Speech enhancement using residual convolutional neural network,''
\newblock in {\em Proc. International Conference on Smart Systems and Inventive Technology ({ICSSIT})}, 2019, pp. 1193--1196.

\bibitem{lstm}
S.~Hochreiter and J.~Schmidhuber,
\newblock ``Long short-term memory,''
\newblock {\em Neural Computation}, vol. 9, no. 8, pp. 1735--1780, 1997.

\bibitem{lstm1}
A.~Mehrish, N.~Majumder, R.~Bharadwaj, R.~Mihalcea, and S.~Poria,
\newblock ``A review of deep learning techniques for speech processing,''
\newblock {\em Information Fusion}, vol. 99, pp. 101869, 2023.

\bibitem{lstm2}
J.~Wang, N.~Saleem, and T.~S. Gunawan,
\newblock ``Towards efficient recurrent architectures: A deep {LSTM} neural network applied to speech enhancement and recognition,''
\newblock {\em Cognitive Computation}, pp. 1221--1236, 2024.

\bibitem{crnn}
Y.~Zhao, X.~Jin, and X.~Hu,
\newblock ``Recurrent convolutional neural network for speech processing,''
\newblock in {\em Proc. {IEEE} International Conference on Acoustics, Speech and Signal Processing. ({ICASSP})}, New Orleans, LA, USA, 2017, pp. 5300--5304.

\bibitem{crnnse}
H.~Zhao, S.~Zarar, I.~Tashev, and C.-H. Lee,
\newblock ``Convolutional-recurrent neural networks for speech enhancement,''
\newblock in {\em Proc. {IEEE} International Conference on Acoustics, Speech and Signal Processing. ({ICASSP})}, Calgary, AB, Canada, 2018, pp. 2401--2405.

\bibitem{tan2018convolutional}
K.~Tan and D.~Wang,
\newblock ``A convolutional recurrent neural network for real-time speech enhancement,''
\newblock in {\em Proc. Interspeech}, 2018, pp. 3229--3233.

\bibitem{hsieh2020wavecrn}
T.-A. Hsieh, H.-M. Wang, X.~Lu, and Y.~Tsao,
\newblock ``{WaveCRN}: An efficient convolutional recurrent neural network for end-to-end speech enhancement,''
\newblock {\em IEEE Signal Processing Letters}, vol. 27, pp. 2149--2153, 2020.

\bibitem{vaswani2017attention}
A.~Vaswani, N.~Shazeer, N.~Parmar, J.~Uszkoreit, L.~Jones, A.~N. Gomez, {\L}.~Kaiser, and I.~Polosukhin,
\newblock ``Attention is all you need,''
\newblock in {\em Proc. Advances in Neural Information Processing Systems}, 2017, pp. 5998--6008.

\bibitem{transformersurvey}
S.~Islam, H.~Elmekki, A.~Elsebai, J.~Bentahar, N.~Drawel, G.~Rjoub, and W.~Pedrycz,
\newblock ``A comprehensive survey on applications of transformers for deep learning tasks,''
\newblock {\em Expert Systems with Applications}, vol. 241, pp. 122666, 2024.

\bibitem{wolf2020transformers}
T.~Wolf, L.~Debut, V.~Sanh, J.~Chaumond, C.~Delangue, A.~Moi, P.~Cistac, T.~Rault, R.~Louf, M.~Funtowicz, et~al.,
\newblock ``Transformers: State-of-the-art natural language processing,''
\newblock in {\em Proc. Conference on Empirical Methods in Natural Language Processing: System Demonstrations}, 2020, pp. 38--45.

\bibitem{baevski2020wav2vec}
A.~Baevski, Y.~Zhou, A.~Mohamed, and M.~Auli,
\newblock ``wav2vec 2.0: A framework for self-supervised learning of speech representations,''
\newblock {\em Advances in Neural Information Processing Systems}, vol. 33, pp. 12449--12460, 2020.

\bibitem{karitaasru}
S.~Karita et~al.,
\newblock ``A comparative study on transformer vs {RNN} in speech applications,''
\newblock in {\em Proc. {IEEE} Automatic Speech Recognition and Understanding Workshop ({ASRU})}, Singapore, 2019, pp. 449--456.

\bibitem{stransformer}
L.~Dong, S.~Xu, and B.~Xu,
\newblock ``Speech-transformer: A no-recurrence sequence-to-sequence model for speech recognition,''
\newblock in {\em Proc. {IEEE} International Conference on Acoustics, Speech and Signal Processing. ({ICASSP})}, 2018.

\bibitem{noise}
G.~Hu,
\newblock ``100 nonspeech environmental sounds,''
\newblock Tech. {R}ep., The Ohio State University, Department of Computer Science and Engineering, 2004.

\bibitem{lstmse}
L.~Sun, J.~Du, L.-R. Dai, and C.-H. Lee,
\newblock ``Multiple-target deep learning for {LSTM}-{RNN} based speech enhancement,''
\newblock in {\em Proc. Hands-free Speech Communications and Microphone Arrays ({HSCMA})}, 2017, pp. 136--140.

\bibitem{timitasr}
L.~Meng, J.~Xu, X.~Tan, J.~Wang, T.~Qin, and B.~Xu,
\newblock ``Mixspeech: Data augmentation for low-resource automatic speech recognition,''
\newblock in {\em Proc. {IEEE} International Conference on Acoustics, Speech and Signal Processing. ({ICASSP})}, Toronto, ON, Canada, 2021, pp. 7008--7012.

\bibitem{timitse}
Y.-J. Lu, C.-Y. Chang, C.~Yu, C.-F. Liu, J.-W. Hung, S.~Watanabe, and Y.~Tsao,
\newblock ``Improving speech enhancement performance by leveraging contextual broad phonetic class information,''
\newblock {\em IEEE/ACM Transactions on Audio, Speech, and Language Processing}, vol. 31, pp. 2738--2750, 2023.

\bibitem{timitse1}
S.-W. Fu, C.-F. Liao, and Y.~Tsao,
\newblock ``Learning with learned loss function: Speech enhancement with {Quality-Net} to improve perceptual evaluation of speech quality,''
\newblock {\em IEEE Signal Processing Letters}, vol. 27, pp. 26--30, 2020.

\bibitem{mosa}
R.~E. Zezario, S.-W. Fu, F.~Chen, C.-S. Fuh, H.-M. Wang, and Y.~Tsao,
\newblock ``Deep learning-based non-intrusive multi-objective speech assessment model with cross-domain features,''
\newblock {\em IEEE/ACM Transactions on Audio, Speech, and Language Processing}, vol. 31, pp. 54--70, 2023.

\bibitem{cedenza2}
{The Cadenza Project},
\newblock ``The cadenza project,'' [Online]. Available: \url{https://cadenzachallenge.org}.

\bibitem{matricse}
S.-W. Fu, C.~Yu, T.~Hsieh, P.~W. Plantinga, M.~Ravanelli, X.~Lu, and Y.~Tsao,
\newblock ``{MetricGAN+}: An improved version of {MetricGAN} for speech enhancement,''
\newblock in {\em Proc. Interspeech}, 2021.

\bibitem{vctkse}
H.~Schr{\"o}ter, A.~N. Escalante-B., T.~Rosenkranz, and A.~Maier,
\newblock ``Deepfilternet: Perceptually motivated real-time speech enhancement,''
\newblock in {\em Proc. Interspeech}, 2023, pp. 2008--2009.

\bibitem{vctkasr}
M.~Ravanelli et~al.,
\newblock ``{SpeechBrain}: A general-purpose speech toolkit,'' 2021,
\newblock arXiv:2106.04624.

\bibitem{musdb1}
S.~Rouard, F.~Massa, and A.~D{\'e}fossez,
\newblock ``Hybrid transformers for music source separation,''
\newblock in {\em Proc. {IEEE} International Conference on Acoustics, Speech and Signal Processing. ({ICASSP})}, Rhodes Island, Greece, 2023, pp. 1--5.

\bibitem{musdb2}
A.~D{\'e}fossez,
\newblock ``Hybrid spectrogram and waveform source separation,''
\newblock {\em arXiv preprint arXiv:2111.03600}, 2021.

\bibitem{hlp}
W.~B. Alshuaib, J.~M. Al-Kandari, and S.~M. Hasan,
\newblock ``Classification of hearing loss,''
\newblock in {\em Update on Hearing Loss}. InTech, 2015.

\bibitem{msbg}
Y.~Nejime and B.~C. Moore,
\newblock ``Simulation of the effect of threshold elevation and loudness recruitment combined with reduced frequency selectivity on the intelligibility of speech in noise,''
\newblock {\em The Journal of the Acoustical Society of America}, vol. 102, no. 1, pp. 603--615, 1997.

\end{thebibliography}

\end{document}